\documentclass[sigconf]{acmart}

\usepackage{booktabs} 
\usepackage{balance}

\usepackage{soul}
\usepackage{xcolor}

\newcommand{\authorcomment}[1]{}

\newcommand {\jingyi}[1]{\authorcomment{\textcolor{purple}{\bf{JL: #1}}\normalfont}}
\newcommand {\sonia}[1]{\authorcomment{\textcolor{orange}{\bf{SH: #1}}\normalfont}}
\newcommand {\changes}[1]{\textcolor{black}{#1}\normalfont}
\usepackage{xspace}
\newcommand{\molly}{Molly\xspace}
\newcommand{\mollyfull}{Molly Morin\xspace}
\newcommand{\lynn}{Lynne\xspace}
\newcommand{\lynnfull}{Lynne Yun\xspace}
\newcommand{\eran}{Eran\xspace}
\newcommand{\eranfull}{Eran Hilleli\xspace}
\newcommand{\miwa}{Miwa\xspace}
\newcommand{\miwafull}{Miwa Matreyek\xspace}
\newcommand{\emily}{Emily\xspace}
\newcommand{\emilyfull}{Emily Gobeille\xspace}
\newcommand{\shantell}{Shantell\xspace}
\newcommand{\shantellfull}{Shantell Martin\xspace}
\newcommand{\michael}{Michael\xspace}
\newcommand{\michaelfull}{Michael Salter\xspace}
\newcommand{\chris}{Chris\xspace}
\newcommand{\chrisfull}{Chris Coleman\xspace}
\newcommand{\fish}{Fish\xspace}
\newcommand{\fishfull}{Fish McGill\xspace}
\newcommand{\ben}{Ben\xspace}
\newcommand{\benfull}{Ben Tritt\xspace}
\newcommand{\nina}{Nina\xspace}
\newcommand{\ninafull}{Nina Wishnok\xspace}
\newcommand{\kim}{Kim\xspace}
\newcommand{\kimfull}{Kim Smith\xspace}
\newcommand{\mac}{Mackenzie\xspace}
\newcommand{\macfull}{Mackenzie Schubert\xspace}

\AtBeginDocument{%
  \providecommand\BibTeX{{%
    \normalfont B\kern-0.5em{\scshape i\kern-0.25em b}\kern-0.8em\TeX}}}

\setcopyright{iw3c2w3}
\copyrightyear{2021} 
\acmYear{2021} 
\acmConference[CHI '21]{CHI Conference on Human Factors in Computing Systems}{May 8--13, 2021}{Yokohama, Japan}
\acmBooktitle{CHI Conference on Human Factors in Computing Systems (CHI '21), May 8--13, 2021, Yokohama, Japan}\acmDOI{10.1145/3411764.3445682}
\acmISBN{978-1-4503-8096-6/21/05}

\begin{document}

\title{What We Can Learn From Visual Artists About Software Development}



\author{Jingyi Li}
\affiliation{%
  \institution{Stanford University}
}
\author{Sonia Hashim}
\affiliation{%
  \institution{University of California, Santa Barbara}
}
\author{Jennifer Jacobs}
\affiliation{\institution{University of California, Santa Barbara}
}

\renewcommand{\shortauthors}{Li et al.}

\begin{abstract}

This paper explores software’s role in visual art production by examining how artists use and develop software. We conducted interviews with professional artists who were collaborating with software developers, learning software development, and building and maintaining software. We found artists were motivated to learn software development for intellectual growth and access to technical communities. Artists valued efficient workflows through skilled manual execution and personal software development, but avoided high-level forms of software automation. Artists identified conflicts between their priorities and those of professional developers and computational art communities, which influenced how they used computational aesthetics in their work. These findings contribute to efforts in systems engineering research to integrate end-user programming and creativity support across software and physical media, suggesting opportunities for artists as collaborators. Artists’ experiences writing software can guide technical implementations of domain-specific representations, and their experiences in interdisciplinary production can aid inclusive community building around computational tools.

\end{abstract}




\maketitle

\section{Introduction}

Software has become closely integrated with visual art production. Illustration, painting, animation, and photography are just a small sample of once non-digital fields where many practitioners now incorporate software. Like all creative tools, software offers artists new opportunities while simultaneously imposing constraints. For example, digital drawing software allows artists to continually edit their pieces through layers and undo, but limits them to the functionality provided on a toolbar and can make it difficult to integrate physical media. 


Developing software for visual art brings many specific challenges. Artists are distinguished from many other software users by their nonconformity and unconventional approaches to making artifacts \cite{feist1999}. Artists also work extensively with physical media and manual forms of expression---qualities that are challenging to integrate in digital representations \cite{schacman10.1145/2384592.2384594, beingmachine-10.1145/2702123.2702547}. Designing software abstractions and constraints for different domains of visual art poses additional challenges. The ways software designers expect artists to interact with the system may be different from the ways artists are used to engaging with materials to make art in their domain \cite{victor2011dynamic}.
Furthermore, artists who use domain-specific creative coding languages \cite{processing, ofbook} face additional challenges in having to understand abstract representations and work in a highly structured manner \cite{victor2011dynamic}. These forms of working can be incompatible with their existing practices of manual manipulation and non-objective exploration \cite{Berger, goodman1968languages}. 

\changes{These challenges, as well the opportunities they surface, affect a variety of stakeholders. They include systems engineering researchers who research creativity support technologies. They also include artists who collaborate with software developers in their own practice. Finally, they include artists who \textit{are} software developers. Such individuals build personal tools, but also can support communities of creative practitioners by leveraging their domain knowledge to develop and distribute new, artist-specific platforms \cite{processing, levin2003essay}}. To support multiple stakeholders in software production for visual art and address these challenges, we sought to develop a more detailed understanding of the ways visual artists work with software. By studying how artists develop and use their own forms of software, we synthesized design implications across the joint interests of artists and professional software developers.

In this paper, we ask two research questions: (1) What factors lead artists to embrace or reject software tools created by others? (2) What factors motivate artists to develop their own software, and how do their approaches diverge or align with other domains of software development? 

Our research questions are motivated by our efforts to develop programming systems for visual artists over the past seven years and oriented towards informing future efforts in systems engineering research for creativity support. 
Like researchers, artists are motivated, in-part, by creating novel outcomes. 
\changes{We argue that artists' motivations for building their own software are well aligned with the kinds of contributions valued by systems engineering for visual art production HCI research.}
Artists' joint concerns around expressiveness, audience, and commercial viability, as well as their unique workflows, can inform approaches in end-user software development. But at the same time, artist communities have different values, norms, and forms of dissemination than those of professional software developers \cite{Golan}. Understanding and integrating these values is the first step in informing research partnerships between these communities.




To investigate these questions and opportunities, we formalized our initial observations into 13 in-depth interviews conducted with artists across a spectrum of software development expertise, deliberately selected from a larger set of conversations over the past four years. Our interviews sought to understand how artists’ tools and materials–-digital, physical, and programmatic--impacted their processes and outcomes. This paper makes two contributions. First, through a thematic analysis of our interviews, we surface themes on how software intersects with visual art practice: how artists were motivated to build software as a mechanism for intellectual and community engagement, how existing software representations worked with or against artists' complex workflows, how artists valued efficiency but resisted forms of software automation, how artists used software beyond making art to organize, motivate and reflect, and how interaction with technical communities impacted artists' aesthetic choices. Second, our findings suggest opportunities for how efforts in end-user programming can map onto creativity support tools for visual artists. 
Current forms of software automation misalign with how artists work, while forms of higher level abstraction hinder their ability to manipulate data representations. We argue that artists, who innovate in adapting and writing software to fit their idiosyncratic working styles, can guide technical implementations of domain-specific program representations and that their experiences in interdisciplinary production can aid inclusive community-building around computational tools. 

\section{Background}
Our research contributes to HCI efforts to inform technical systems through the study of creative communities. Because we studied how visual artists develop software, our work builds on and informs end-user programming research. To guide our analysis, we also examined artist-built programming tools and communities, and research-driven creativity support tools.

\subsection{Informing HCI by Studying Creative Practice}
Understanding creative practice requires different forms of inquiry beyond controlled study~\cite{Shneiderman_2007}. By investigating art and craft practice, researchers have demonstrated alternative strategies for established domains of HCI. This includes studying ceramics to inform interaction design~\cite{Bardzell_Rosner_Bardzell_2012}, furniture production to inform CAD and digital fabrication research~\cite{Cheatle_Jackson_2015}, and house construction to inform design for living materials~\cite{Dew_Rosner_2018}. Researchers have also used collaborative models working with artists or craftspeople to engage in joint production~\cite{Jacobs_Zoran_2015} or co-author HCI publications~\cite{Devendorf_Arquilla_Wirtanen_Anderson_Frost_2020,Fiebrink_Sonami}. These investigations challenge notions that artists or craftspeople have limited technical proficiency, demonstrate the technical sophistication of their work, and describe concrete ways in which they can inform technical systems. Inspired by prior efforts, we focus on how software use and development by visual artists can inform end-user programming and systems engineering research. 

Studies of art practices have also examined the role of manual production in creative workflows. HCI researchers have theorized that ideas emerge from interacting with materials \cite{Ingold_2010, Schon}, tangible making builds mental models 
\cite{Papert-mindstorms, Tversky}, and physical manipulation facilitates concrete cognitive tasks 
\cite{Gross, Klemmer_Hartmann_Takayama_2006}. 
Manipulating physical media helps artists develop manual skill and knowledge \cite{Needleman}, react and plan in step \cite{Berger,Do_Gross}, and reason about building tools \cite{McCullough}. Manual tools also afford expressiveness, by preserving gesture \cite{Mumford}, and speed, which supports open-ended exploration \cite{Berger}. In contrast to prior work on manual production, we investigate how efficiency and automation in software align or conflict with artists’ manual processes and aesthetic preferences. 

\subsection{Creative Coding Systems: Programming Tools \& Communities for Visual Art}
The expressive power of programming has led visual artists to develop and disseminate their own programming tools. Lieberman, Watson, and Castro developed OpenFrameworks, a textual programming toolkit, to create interactive audiovisual artwork \cite{ofbook}. The textual languages Processing and p5.js \cite{processing_overview, p5.js} emerged from exploring how to teach programming through the lens of art \cite{processing_overview}. Node-based visual programming frameworks like Max and vvvv apply ideas from signal processing for computational music towards producing computational artwork \cite{max, vvvv}. 
\changes{
Researchers in music technology and live-coding have developed domain-specific representations and tools \cite{Eaglestone2001, Barbosa2018, WorkbenchMusic2019}. The goals of these domains differ from visual art. For instance, latency and scrubbing metaphors are less relevant to visual art. Research on high ceilings in computational music achieved via design metaphors of control intimacy of musical instruments \cite{Wessel2001} is different from our work that also focuses on expert practitioners. Our focus is to study the ways visual artists build software as a means to inform creativity support systems research. 
}



Some artists who create software tools also engage in community building. 
From the OpenFrameworks design philosophy that prioritizes a ``do it with others'' approach to making art \cite{ofbook}, to the Processing community that builds and maintains the platform's many extensions \cite{processing_overview}, to the p5.js community statement that recognizes diverse contributions from new programmers \cite{p5.js-community}, all of these frameworks rely on collaborative development from their communities. 
The School for Poetic Computation (SFPC), a school run by artists for artists who often build computational tools, is another example of an artist-led creative coding community that extends software use \cite{jacobs-sfpc}. Artist-developed software tools align with art practice and are shaped by community engagement. Our work provides greater detail on this process by examining how artists move from creating artifacts to authoring software and how artists' software use is shaped by interactions with technical communities. 

\subsection{End-User Programming in Art and Design}
Research in end-user programming (EUP) supports non-professional programmers as they develop software to further their work~\cite{lieberman2006end} and has focused largely on interaction designers.  
Research has shown that visual designers seek programming tools that directly integrate with visual drawing tools \cite{myers2008designers} and use high-level tools mapped to specific tasks or glued with general purpose languages rather than learn new programming frameworks \cite{Brandt_Guo_Lewenstein_Klemmer_2008}. 
Systems like Juxtapose \cite{Hartmann_Yu_Allison_Yang_Klemmer_2008} and Interstate \cite{Oney_Myers_Brandt_2014} improve programming for interaction designers through better version management and visualizations. Re-envisioning software as information substrates \cite{Beaudouin-Lafon_2017} that integrate data and application functionality supports greater software malleability and more varied forms of collaboration in web \cite{Klokmose_Eagan_Baader_Mackay_Beaudouin-Lafon_2015} and video editing \cite{Klokmose_Remy_Kristensen_Bagge_Beaudouin-Lafon_Mackay_2019}.

While there has been extensive EUP research targeting designers, less research examines EUP for visual art. 
Researchers have developed a variety of graphic art tools that enable programming through direct manipulation. Some systems support two pane interfaces that place visual output side-by-side with code \cite{Chugh-2980983.2908103, McDirmid-apx}. Recent work demonstrated that allowing users to directly manipulate intermediate execution values, in addition to output, minimized textual editing \cite{Hempel-3332165.3347925}. 
Other work, like Dynamic Brushes, aims to support expressiveness through a direct manipulation environment coupled with a programming language 
\cite{Jacobs_Brandt_Mech_Resnick_2018}. Results from a study of debugging tools developed for Dynamic Brushes suggested that artists inspect program functionality while making artwork \cite{Li-ddb}. Our research is aimed at informing future efforts in EUP for visual art by investigating how artists approach software development and work with software representations. We provide insights into the ways that visual artists' objectives differ from other end-user programmers and highlight opportunities in building domain-specific programming tools for visual art.

\subsection{Creativity Support Tools for Visual Art}

Creativity support tools researchers have extensively studied how software can support visual art workflows. 
While Shneiderman outlined several opportunities for creativity support including exploratory discovery and record keeping \cite{Shneiderman-csts},  many HCI systems for visual art emphasize producing artifacts. Systems such as large-scale generative design visualizations \cite{Matejka_Glueck_Bradner_Hashemi_Grossman_Fitzmaurice_2018, Zaman_Stuerzlinger_Neugebauer_Woodbury_Elkhaldi_Shireen_Terry_2015}, cross-modal generative sketching for 3D design \cite{Kazi-dreamsketch}, or text-based icon design \cite{Zhao-iconate} aid practitioners in exploring ideas and selecting  artifacts. Researchers have also explored supporting specific affordances in tools such as speed-aware line smoothing algorithms \cite{Thiel-elasticurves}, manipulations of negative space to edit vector graphics \cite{Bernstein-lillicon}, or selective undo in digital painting \cite{Myers-undo}. Another category of systems examine ways to digitally emulate physical forms of production \cite{Barnes-video-puppetry, Kazi-sandcanvas, Leung-greasepencil}. 

A large body of creativity support research focuses on broadening participation. Shneiderman, as well as Silverman and Resnick \cite{Resnick_Silverman_2005}, advocated for creative systems with ``low-floors'' that reduce the barriers to entry and ``wide walls'' that support a diverse range of outcomes. Researchers have examined reducing barriers to making computational art through direct manipulation interfaces for creating procedural constraints \cite{Jacobs-para,Kazi-kitty}. Other systems guide novices in tasks like photographic alignment \cite{E-photography} and narrative video composition \cite{Kim-motif}. Machine learning (ML) based systems including applications of neural style transfer (NST) \cite{Gatys, Champandard, Iizuka}, user-in-the-loop tools \cite{RunwayML}, or support for specific automated tasks like auto-complete for hand-drawn animation \cite{Xing-autocomplete-anim}, sketch simplification \cite{Simo-Serra-2}, and layout generation \cite{Batra-layout}, are increasingly used to facilitate easy entry to visual art production through high-level automation. ML-based tools have raised new questions about the relationships between artists and software. Semmo et al. challenged the use of NST arguing that NST must support interactivity and customization for use in creative production \cite{Semmo}. Hertzmann critiqued the notion that ML-based automation creates AI-generated art, arguing that computational systems are not social agents and therefore cannot make art \cite{Hertzmann}. We seek to understand and critique high-level forms of software automation for visual art by examining the ways artists use or reject these systems in practice.

\section{Methods}
This work is structured around a thematic analysis of interviews with professional visual artists. These interviews were motivated and informed by the authors' personal experiences working between systems engineering and fine art. 

\subsection{Author Backgrounds}
To understand the perspectives that shaped our work, we provide background on the research team's expertise and focus. The authors represent a spectrum of art experience: Jingyi maintains a hobbyist painting and illustration practice, Sonia studied art history, and Jennifer has formal art training and worked as a professional artist. Presently, Jingyi and Sonia are graduate students in computer science and Jennifer is an HCI professor in an interdisciplinary art and engineering department. As we transitioned from creating artwork to researching and building software tools, we had the opportunity to test our tools with practicing artists. Our discussions about software use, programming, and creative production with artists indicated differences in how practicing artists and professional software developers viewed the opportunities of software.

\begin{table*}

\begin{tabular}{|l|l|l|l|l|}
\hline
\textbf{Name}  & \textbf{Years Active}  & \textbf{Description} & \textbf{Role}  \\ \hline
\mollyfull  &  10+     & Digital fabrication artist  &  Studio artist, fine arts professor                           \\ \hline
\lynnfull  &  15+   &  Letterform designer   &  Designer at type firm           \\ \hline
\eranfull  &  10+  & Animator & Animator, software developer                             \\ \hline
\miwafull  &   10+   & Animator \& dancer &  Independent artist       \\ \hline
\emilyfull   &   15+     &  Interaction designer \& illustrator    &   Studio founder                        \\ \hline
\shantellfull & 15+ & Large format \& visual artist & Independent artist                                                       \\ \hline
\michaelfull &  25+  & Studio artist  &  Studio artist, fine arts professor    \\ \hline
\chrisfull & 15+ &  Emergent media practitioner &  Studio artist, fine arts professor     \\ \hline
\fishfull &  8  & Pen \& ink illustrator  &  Studio artist, fine arts professor    \\ \hline
\benfull  &  20+  & Painter & Studio founder    \\ \hline
\ninafull &  10+  & Printmaker &  Studio founder   \\ \hline
\kimfull &  10+  & Painter \& illustrator &  Studio artist, art education company founder    \\ \hline
\macfull & 10+  & Illustrator  &  Studio artist, art technology company founder  \\ \hline

\end{tabular}

\caption{Artist demographics.}
\label{table:participants} 
\vspace{-20pt}
\end{table*}
\subsection{Interview Methodology and Participants}
Building from this preliminary observation, we conducted a formal set of semi-structured interviews with 20 professional visual artists over a period of four years. We recruited subjects through our established networks drawn from the artist residencies, exhibitions, and educational networks that \changes{Jennifer} participated in. We aimed to interview artists who worked across a diverse set of materials (e.g., software, paints, code), processes (e.g., coding, manual drawing, performance), and domains (e.g., illustration, animation, interactive art). Figure \ref{fig:art} shows sample artworks from each artist and Table \ref{table:participants} shows basic demographics information, \changes{with full artist names released with consent}. 
While these interviews were interspersed with and helped guide research in building software tools, in this work, we foreground the insights from these interviews, rather than presenting a limited subset to motivate a specific system.

For this paper, we included data from 13 out of the 20 interviews. We omitted interviews from artists who did not work with software or visual art as well as those in which the artist focused on their conceptual stances over their process. Our interviews were primarily in person, with five conducted through video conferencing software, and on average lasted 1.5 hours. Our interview objectives were to understand what artists perceive to be the primary opportunities and limitations of digital, manual, and physical media; how different media shaped artists' learning, process, and outcomes; and what factors encouraged or prevented artists from engaging in software development in their work. Prior to each interview, we reviewed each artist's work to direct process-based questions towards specific pieces.

\subsection{Data Collection and Analysis}
We audio recorded and transcribed each interview. For analysis, we conducted a reflexive thematic analysis~\cite{braun,braun_reflexive}, focusing on an inductive approach open to latent themes. Each author reviewed each transcript, and, following a discussion of initial patterns, each author coded a subset of transcripts to initially identify as many interesting data extracts as possible. The research team refined the codes and conceptualized them into preliminary themes through weekly meetings and discussions. After all authors collaboratively drafted a written description of each theme, \changes{Sonia and Jennifer} reviewed them with respect to the coded transcripts to confirm they were representative of the original codes.

\subsection{Limitations}
Our work focuses on a deep qualitative examination of 13 artists. This approach was necessary to gain insight into the specific factors that shaped the workflows of our interviewees. \changes{Artists represented here had a range of experiences with software; future research engaging exclusively with artists who build software will likely uncover additional details such as a quantitative breakdown of time spent across different software development tasks.} We chose our methodology based on our personal experiences transitioning between art and software development and research, which we disclose in our author background statement to contextualize our analysis and discussion. Any interview risks social desirability bias. Because we had personal connections with our interviewees, trust contributed to them being comfortable giving honest answers, increasing the reliability of the responses \cite{fujii2017interviewing}.

\section{Intersections of Software and Visual Art}
We conceptualized themes across six dimensions presented below. Revisiting our research questions, we describe how software constraints and representations, use cases beyond producing art, and cultural associations of computational aesthetics impacted how artists used software tools made by others. \changes{Artists were motivated to develop their own software to create new functionality in their works, grow intellectually, and gain technical legitimacy in their communities. Furthermore, their idiosyncratic workflows spanning digital and physical mediums were often misaligned with both the constraints and forms of automation provided by existing software.}

\subsection{Artists' Motivations for Software Development}\label{motivations}
 \begin{figure*}[t]
\includegraphics[width=\textwidth]{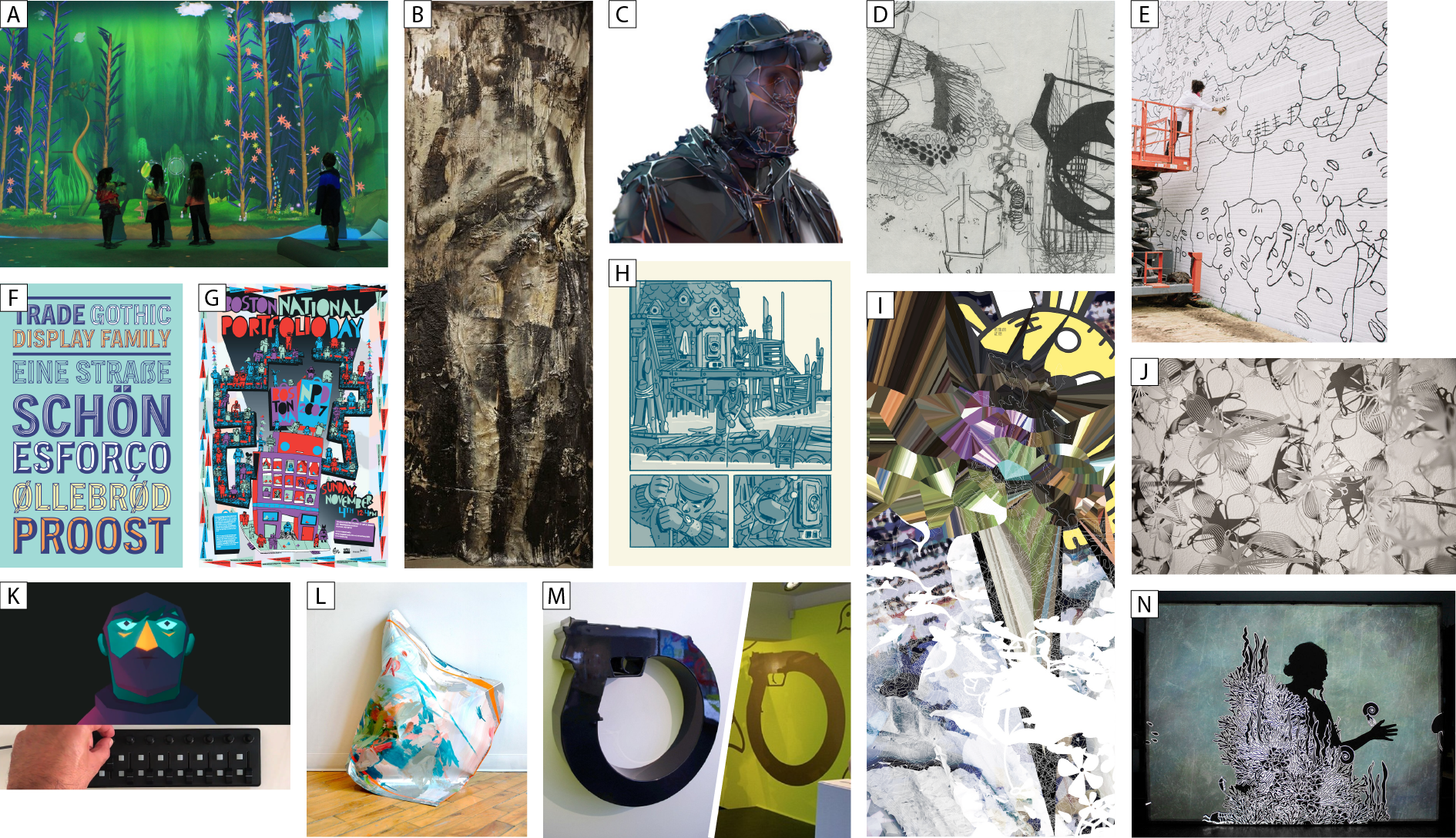} 
 	 \caption{Artwork created by interviewed artists: A) Interactive installation by \emilyfull. B) Oil painting by \benfull. C) New media sculpture by \chrisfull. D) Pencil illustration by \ninafull. E) Line drawing mural by \shantellfull. F) Typeface by \lynnfull. G) Poster by \fishfull. H) Comic page by \macfull. I) Digital illustration by \michaelfull in collaboration with \chrisfull. J) Vinyl cut sculptures by \mollyfull. K) Synthesizer controlled animated character by \eranfull. L) Painting by \kimfull. M) 3D printed vector graphic by \michaelfull. N) Shadow installation by \miwafull.}
\Description{14 pieces of artwork arranged in a grid. A: People interacting with a wall with a projected green background and plant-like structures. B: A textured monochromatic brown oil painting of a human figure. C: A digital sculpture of a person wearing a hat with web-like textures. D : An abstract pencil line drawing on light paper, with many geometric forms and an abstract house. E: An artist standing on a lift drawing a large line drawing mural on a wall. The mural has a white background and is of people’s faces. F: A poster demonstrating a font called “Trade Gothic Display Family” in all uppercase, sans-serif, bold lettering. G: A poster advertising Boston National Portfolio Day, using primarily purple, red, and blue, with many stylized people walking into a building. H: A three panel digital comic book page done entirely in shades of blue on a yellow background. A man is outside of a cluttered house pulling up something from a cable. I: A digital piece that morphs hand drawn illustrations with procedurally generated forms. The top half of the piece is various yellow lines with an indistinguishable vector graphic and the bottom half is white with some flowers. J: Vinyl cut sculptures, thin white floral shapes, hanging on the wall with shadows accentuating their form. K: On the top, a 3D modeled man with a beard, blue skin, and a yellow nose. On the bottom, the artist interacting with a synthesizer to control the man. L: An abstract painting, mainly greys, blues, and oranges, on a large canvas that has been deliberately crumpled and reshaped. M: A black 3D printed gun, except the barrel of the gun loops in a circle to the mouth of the gun. N: A wall showing a projection and the shadow of a human. The background is teal and there is a black and white anemone form obscuring part of the human shadow.}
      	\label{fig:art}
 	\vspace{-10pt}
\end{figure*}

Out of the 13 artists from our interviews, eight either were software developers or were in the process of learning software development. We distinguish between software development and programming in that programming involves writing code, while software development also encompasses testing, maintaining, and sharing software \cite{Ko_Abraham}. Our software developers described a variety of motivating factors. Initially, artists developed software to add interactivity to their artwork. 
For example, \mac learned Unity development to facilitate dynamic 3D transitions between panels in an interactive digital comic, and \emily learned C++ and Macromedia Director's Lingo programming language because it enabled her to create interactive animations. \changes{When adding software as a component of their output, artists valued robustness.} Artists also developed software to automate tasks.  \changes{While using software as a tool for art-making, artists placed a higher emphasis on reusability.} 
\molly~learned to procedurally generate vector graphics to reduce manual labor when repeating forms in Adobe Illustrator, and \lynn~learned Python to reduce the effort required to test different combinations of type when creating new fonts. These objectives align with the established framing of end-user programmers as individuals who write code in service of another practice. 

In addition to functional applications, artists developed software because of the opportunities for intellectual and creative growth. For instance, \ben described an interest diving into programming in order to ``look under the hood'' of software. Redesigning or building new software systems enabled artists to identify constraints in their tools to imagine alternatives. \lynn's initial experience with Python acted as the catalyst for her to enroll in a creative programming course. She described how learning to use a C++ based framework to make her own graphical user interfaces (GUIs) enabled her to recognize third party software constraints and envision other options: 
\begin{quote}
   \textit{If I could make my own GUI for things, maybe I could be using things in a different manner. This is a recent thought\dots I don't think I ever realized how much it was impacting my work.}
\end{quote}
The notion that authoring software could expand one's awareness of creative possibilities was remarkably consistent among the artists we spoke with, though artists varied how they used this idea in practice. \emily, \chris, and \eran all developed their own software interfaces that exposed specific parameters to explore and fine-tune visual properties of their work. \molly~ described the intellectual satisfaction she derived from translating her manual process generating vector geometry for fabrication to an algorithmic description, enjoying solving complex geometric problems while creating a reusable tool that reflected her manual practice. 
Similarly, \eran created experimental software tools primarily to investigate new concepts, and he was less concerned if the resulting tool would lead to a finished piece. \changes{Artists instead valued speed in designing software sketches to quickly test ideas.}

Finally, four artists described being motivated to develop software to influence future forms of software design, different from what they currently observed. \eran and \chris supported students and newcomers to animation and embedded programming, respectively, by designing tools that addressed obstacles they experienced in their own work. \lynn described how developing software would allow her to ``have a seat at the table'' around media software production. 
Likewise, \chris recognized that, as an independent artist and professor, he couldn't compete with the speed and resources of professional software companies, but he could release different kinds of tools that influenced the direction those companies might take, saying, ``[a]ll I can do is shape the conversation for the professional tools that get made afterwards.'' 
\kim went a step further, describing her desire to improve her software development skills as a means to bypass negative experiences and communication breakdowns she had experienced when working directly with professional developers: \begin{quote}\textit{It's not always clear to the developers that I've worked with why things are important from a designer's point of view \dots I think if I could really get that skill down and design as well \dots the end results would be better.}\end{quote} 

These experiences demonstrate how learning and participating in software development enabled multiple forms of power in visual art production. First, artists developed software to create new functionality in artworks. Second, artists developed software to grow intellectually and they built their own interfaces to explore or refine work. Finally, by demonstrating knowledge in software development, artists gained technical legitimacy and could engage in dialogue with professional software developers or circumvent them altogether.

\subsection{Selecting Software Constraints and Representations}\label{software}
Our interviews revealed two ways software made by others impacted artists' processes and outputs: they negotiated different forms of software constraints, and they carefully selected specific graphic representations. 

Artists viewed software constraints as constructive when they could define the constraints' parameters. Sometimes, this involved choosing to not use a tool's features. \shantell, \eran, \fish, \ben, and \mac all described points when they constrained themselves from using software-based undo. This constraint replicated the quality of physical ink, forcing them to work with their mistakes or avoided breaking their flow of drawing with editing. Artists also imposed constraints on their practices through their formal knowledge of design and composition. \michael described how, in Illustrator, he manually laid out his compositions to follow grid structures but broke those structures at arbitrary points---a process that would have been more difficult if Illustrator enforced the grid constraint. In other instances, artists developed software to enact constraints. \emily, \eran, and \fish authored software tools that restricted a user's ability to erase, define geometry, or select colors. \fish described programming a drawing tool that automatically faded past strokes over time: 
\begin{quote}
    \textit{I worked in Processing on creating drawing tools that would fade over time\dots I was storing screenshots of every stroke, so then I could watch how someone's drawing came together on a loop\dots and that came out of experiences, just looking for ways to get other people that are afraid of drawing, just to jump in and try something. So, it was a combination of creating the software that would capture each line and mark and letting it fade, so they can get a sense of depth as they are working.}
\end{quote}
While artists had positive experiences imposing their own constraints, they struggled with the stylistic constraints imposed by feature-rich commercial software tools. Despite their respect for and reliance on commercial tools like Adobe After Effects and Unity, \miwa, \molly, \nina, \eran, and \chris struggled with feeling like stylistic aspects of their work were, in \molly's words, ``predetermined by the program.'' In part, this reaction was tied to the expectation for fine artists to create novel imagery. \miwa and \chris both described their need to obscure the means of development when using After Effects. The constraints of high-level software tools were also at odds with artists' desires to enact custom workflows. Artists avoided ``defaults'' and ``presets'' for this reason. For example, \nina avoided Photoshop filters because they were ``straight out of the box'' and incompatible with her personal workflow, and \fish described feeling ``stifled'' by the aesthetic constraints of Illustrator defaults until he learned how to author custom brushes.

Similar to constraints, artists embraced or rejected lower-level graphic representations---e.g., 3D meshes, Bezier curves, bitmaps---based on the extent that a representation supported their workflow.
Both \michael and \molly worked primarily with vector graphics, despite being adept in other representations, because vectors were best suited to the curvilinear geometry and ``clean'' aesthetics of their work. 
Similarly, other artists deliberately rejected some digital graphic representations or were frustrated with the inconsistencies that emerged when they tried to blend two different representations. \emily described an extreme dislike for the ``smoothness'' of 3D graphics, and \mac described how integrating work from Photoshop and Illustrator created a stylistic ``gap.'' In most cases, these tensions with software representations were not the result of nostalgia for physical media. \emily wasn't interested in recreating traditional techniques in a digital format, stating her goal was to ``push both traditional artistic practice and software tools out of their individual comfort zones, to create something that is unique and blurs the lines between the two.'' Alternatively, \shantell, \michael, and \mac felt digital representations actually had comparable aesthetic qualities to physical media. 

Instead, the degree to which artists were comfortable working with a digital representation was determined by how it supported their workflows. 
At a low level of individual mark making, \mac described how Bezier curves afforded editing existing work (a process he described as ``finicky'') whereas bitmap brushes pushed him to ``[sketch] one thing once and move through it,'' because the bitmap representation didn't support the same level of editing after drawing. At a higher level, \eran described how the timeline representation in animation software required animators to painstakingly edit individual frames. As an animation instructor, he observed students transitioning between drawing the animation and adjusting the timeline to the point of fatigue:
\begin{quote}\textit{The task the person has to do is finish his drawing, move the timeline, [and] change something. I see students that, every time they do that, their mind hurts\dots It's a break in their flow.}\end{quote} 
These observations led \eran to develop animation tools with continuous, rather than frame-based, timelines. He recognized that these different representations lead to trade-offs in workflows---a continuous timeline would afford speed and shift the focus to drawing, but a key-frame timeline enabled low-level but laborious control. 

Artists also considered aesthetics clearly shaped by a specific representation to be an indicator of novice work. \michael and \molly described how, as teachers, they worked with students to master software so that their own drawing style was preserved, rather than obscured by the qualities of vector graphics. \mac 
described how representing applications as standalone packages restricted iteration and workflows across different tools:
\begin{quote} \textit{In some ways those programs are really behind kind of walls and are not very modular. The program is this just inside of this} [indicating to software window] \textit{and it has many little tools that exist inside of that and some of those have like smaller tools that exist inside of that. So you're kind of using like all of these tools in Photoshop and all of these tools in Illustrator\dots but when I think about the way I work\dots passing things back and forth and iterating in different ways and quickly. I feel like you've got these big walls between programs\dots I'm interested in\dots how those things can be broken out or built out separately.}
\end{quote}

Overall, the degree to which artists embraced software constraints and representations corresponded to the degree that they could be adapted to an established workflow. When faced with complex tools with powerful black-box functionality or high-level representations, artists often tried to use these tools for unintended purposes. When this was not feasible, they opted for software tools with limited functionality or built their own.



\subsection{Non-Linear Physical-Digital Workflows}\label{physdigworkflows} 

Despite working across a wide range of visual art domains, each artist described workflows that integrated digital and physical processes, working non-linearly between digital and physical production using a diverse set of tools and approaches. Foremost, artists heavily relied on the ability to provide manual inputs to digital tools. \miwa, \ben, and \nina all mentioned they liked the ``organic'' and ``warm'' quality of hand drawn art---deviations and irregularities in their artistic style that were built up over learning to draw and physically engaging with their body. Artists also improvised physical materials as digital inputs. \emily photographed and scanned objects to turn into textures to ``incorporate as many non-computer elements into the digital artwork,'' 
while \nina traced over copies of architectural plans as a starting point for her prints.

Similarly, artists produced physical outputs with digital fabrication. 
\shantell used a CNC mill to fabricate a previously ink-based drawing as functional printed circuit board traces. \michael described how he would arbitrarily decide to convert flat vector graphics to 3D physical objects without advanced planning (Figure \ref{fig:art}M):
\begin{quote} \textit{I can extrude this. I can laser cut it 15 times and laminate it. \dots There's something to me, the formal experience of taking something that changes dimension, which is exciting. I'm gonna find something that I normally couldn't have\dots Those opportunities past the computer.} \end{quote} 


Artists also used physical production as a means to think through problems, either individually or with collaborators. 
For example, \emily worked out the computational rules to define an interactive generative puzzle game while working on a hand-sketched maze. Likewise, \molly shared how she would ``solve code problems'' while felting, constantly ``switching back and forth between doing a little coding and doing a little felting or folding.'' 
\emily and \michael, two artists who used software tools developed by their collaborators, shared that they would evaluate the tools by iteratively building physical artifacts with them. Both sets of collaborators moved back and forth between digital and physical production to create artwork together. For example, in his collaboration with \chris, \michael would ``manually rip'' outputs enabled by \chris's tools and ``start to draw with'' them---he then sent these drawings and feedback to \chris, who in turn would modify his code. 

Moving between digital and physical spaces enabled artists to leverage the affordances that emerged from using both spaces when producing work, such as for painters \ben and \kim. \ben described working out ideas by alternating between digital painting on a tablet, where he could use color pickers to explore color choices instead of manually mixing paints, and explore material considerations through physical painting. He said the materiality of these physical paints added an abstraction to the way a piece communicated an idea beyond its literal representation, something that wasn't available in digital software. Likewise, to inform her physical paintings on canvas (Figure \ref{fig:art}L), \kim relied on ``huge [digital] libraries of images and washes'' to develop her ideas. For drawings to ``look markedly better and more human,'' \fish encouraged his students to cycle through physical and digital drawing:
\begin{quote}
    \textit{Why don't you put a piece of tracing paper over the screen right now, and just physically draw it? And then let's look at the drawing. And then let's open up Illustrator, and then create a version of it again.}
\end{quote}


While appreciative of the benefits software tools provided, artists also encountered challenges of scale when translating physical artifacts to digital tools. \emily, who always began her process with physical sketches, described digitizing, segmenting, and sharing her sketches as unnecessarily laborious and repetitive. Artists were also unable to preserve the ways they manipulated physical elements of an artwork to explore scale and composition. Both \miwa and \emily noted the difficulty of using a screen to design animations that would be projected at large scale (Figure \ref{fig:art} A \& N). 
Likewise, because she determined the scale of her paintings relative to her body, \kim was unable to make the same judgments while working with software interfaces. \nina went ``back and forth'' from working in software to printing out and ``literally'' putting down work on the floor to look at it when making judgments about layout and composition. Both \lynn and \nina elaborated how ``proportions and visual relationships'' needed to be assessed physically because balance and weight were perceived differently on a screen.  

In summary, artists flexibly and non-linearly moved back and forth between physical and digital spaces when creating work. Artists relied on physical manipulation as a means to refine producing artwork and as a form of reflection or problem solving. Finally, they encountered barriers when they were unable to use physical manipulation or embodied notions of proportion and scale in software tools.

\subsection{Valuing Efficiency and Resisting Software Automation}\label{efficiency}

We observed that artists cared deeply about efficiency in their practice; some even developed their own forms of automation. While quickly working \textit{manually} was important for aesthetic outcomes, existing forms of software-enabled automation imposed undesirable aesthetics that artists had to go back and manually refine. 

Artists valued speed and efficiency particularly in contexts of idea exploration, iteration, and turn-around time when working with collaborators. For example, \emily wrote software to procedurally generate and explore many different compositions, as this was faster and less effort than manually creating each one. 
\eran built tools for new ways of artistic expression with the goal of ``getting to the quickest way [he] could test'' them. In his collaboration with \michael, \chris described the importance of speed:
\begin{quote}
 \textit{
 I love the fact that I have just enough proficiency with Processing that in a day we could produce 20 different interesting iterations and then have a longer dialogue about successes and failures and ways to change and ways to improve.}
\end{quote}

In forms of manual art production, working efficiently also resulted in desired aesthetic outcomes. For example, for \shantell, \mac, and \michael, \changes{speed was synonymous with confidence in drawing and crucial to the aesthetic development of their line. }
In contrast, we saw artists reject aspects of automated efficiency that led to undesirable aesthetics, especially when they already possessed the manual skills to do something that looked better than what the software could. For example, because she disliked how the default algorithm on the vinyl cutter produced intersections, \molly wrote her own to outline Bezier curves with some thickness in order to vinyl cut her line drawings (Figure \ref{fig:art}J). Similarly, \lynn hesitated to ever arrange typography along a path because she disliked how the automated result looked:
\begin{quote}
\textit{In Illustrator\dots if you try to set text on a path in that circle, it looks really crappy, so I'll never do it. But in an analog format, where I can cut and paste the letters or draw them to be there, it looks fine\dots Maybe it's because of the program that it was almost taboo for me to put type on a shape, because it looked terrible in the interface.}
\end{quote}


In fact, artists often chose to create works by hand even when they recognized code could have achieved a similar aesthetic outcome. For instance, \michael preferred to execute a painting that had generative art aesthetics by hand because, to him, drawing was more efficient than the overhead of programming a similar result. 
 
Many forms of software-enabled automation that artists relied on were established features, such as undo, redo, layers, saving multiple versions of a file, and digital editability. Artists described liking these forms of automation since they remained ``in the loop'' and still had aesthetic control over their pieces. Taken together, the experiences of artists using software for automation were at odds with the notion that automation would remove tedious manual labor. On the contrary, because artists lacked control over nuanced outcomes in automated systems, they often spent time fighting to achieve their desired aesthetics. \changes{Manually executing their pieces, on the other hand, was both expressive and efficient.} 

\subsection{Using Software Tools Beyond Producing Artifacts}\label{beyond-artifacts}
Software tools served artists in aspects of their practice beyond working on a visual artifact. In this section, we report on visual artists' experiences of using software tools across tasks of documenting, tracking, generating, and sharing ideas, as well as reflecting on the process of drawing.

Artists described using digital software to collect, organize, and refer to artifacts, including sketches from their processes, while making artwork. \ben described storing and relying on digital recipes of how to mix precise colors of paint. \fish would often have his sketches in an art board to the side while working on a main piece, saying it was like having a ``life raft'' to have ``some kind of composition to play off of.'' \miwa shared a similar process, using Evernote to organize and reference inspiration she came across while working and her own drawn notes. \mac depended on both Evernote and rigorous commenting in his code as a means to quickly resume personal projects after working on client work. He described his C\# code in Unity as being ``half comments, at least.'' 

Artists also identified points where their software tools for organization fell flat. \emily, who frequently blended drawing and note taking, felt like compared to a paper notebook, using a computer was ``less freeform.'' \nina felt like duplicating her Illustrator artboards as to not lose old iterations while managing version history ``wasn't streamlined'' and her ideas were ``getting muddled,'' as the cumbersome versioning obscured her creative decision making.  


To aid in reflecting on and analyzing their own processes, artists used forms of digital software, such as video recording. In creating a projection-based animation piece (Figure \ref{fig:art}N), \miwa recorded footage of herself moving around the space to analyze how her shadow affected the piece and refine its composition. Likewise, \fish recorded himself drawing with a webcam in order to review and reflect upon his process ``like a sports commentator.'' \shantell also described recording herself, to rather simulate the pressures of having a live audience, which mentally forced her to draw. Beyond video, \shantell also cared about analyzing the metrics of her artwork and speculated on using computational tools to aid in understanding things she could not see: 
\begin{quote}
\textit{I drew [Figure \ref{fig:art}E] at an average of five inches per second and I had someone work out the combined amount of line---it's roughly 1668 feet long. Oh, and the average coverage is between 10\% and 12\%. So, now what can I do with that information? One thing I'm interested in as an artist is, can I break down all the analytics of my drawing? Can I break down my speeds, my angles, my distances?
}
\end{quote}

A few artists built their own software tools to aid in reflection. The Processing extension \fish created to observe how other artists built up their drawings (as quoted in section \ref{software}) was originally for his own practice, but he also found it valuable as a teaching tool for his students. \nina, on the other hand, was not interested in using tools for analyzing her own workflows because she felt like her lack of experience with coding prevented her from conceptualizing how those tools would work and how she would apply them to her own practice.

In summary, artists engaged with software not only to create visual artifacts but also to organize and share materials in support of their pieces, as well as to introspect on their own and others' artwork. The fact that artists prioritized using computational tools for reflection, analysis, organization, and management showed how they leveraged computational affordances to assist creative labor, as opposed to performing it. In our interviews, artists discussed at length computational tools that helped them in their process of making artwork, as opposed to tools that made the artwork itself.   

\subsection{Relationships between Aesthetics and Audiences}\label{aesthetics}
As previously discussed, software shaped the visual characteristics of artists' work. Artists' decisions to obscure or embrace computational aesthetics were impacted by social and cultural perceptions of technology. Moreover, mismatches between their own values and those of established technical communities initially led artists to hesitate in identifying as technical creators.

Many artists felt artwork produced with generative algorithms trended towards a specific aesthetic, referring to work that was ``generative'' or ``glitch''-based in style, and work that deliberately suggested a technical sophistication by emphasising ``shiny,'' ``sexy,'' or ``hardcore'' elements in its construction. This idea is consistent with discussions in the computational art community around a digital or generative art ``vernacular'' of established and sometimes cliché aesthetics from a narrow set of algorithms~\cite{watz-eyeo}. 

\chris, \michael, \miwa, \nina and \molly described how their audiences' expectations surrounding computationally-produced artwork determined the ways in which they obscured or emphasized these aesthetics. 
In designing computationally generated or interactive works, \chris, \nina, and \miwa all tried to highlight the concepts in their pieces that were \textit{not} about technology. 
When artists chose to incorporate a recently developed computational technique into their work, the incentive for originality and novelty created a contest to quickly map out all possible variations or unconventional applications. Citing the example of Google Deep Dream \cite{mordvintsev2015deepdream}, \chris described how:
\begin{quote} \textit{There's this weird race to find the new edges of the new box every time an update\dots or a new platform is pushed out because you know all the easy stuff is going to be consumed into a more easy popular culture. Constantly finding new aesthetics or what are the aesthetics that everybody is sort of working in, but that you need to push just beyond.}
\end{quote}
\michael, \chris's frequent collaborator, described his appreciation for \chris's ability to ``finesse'' the high-tech components of the work so that they were ``embedded, subdued, and poetic,'' 
recognizing that investing in a high-tech digital aesthetic required artists to continually adapt to rapidly changing trends and avoid clich\'es. He personally chose to avoid this ``baggage'' 
in his own practice. 

When artists brought computational work into more traditional art communities, they had to decide between obscuring the computational qualities of their work or devoting significant effort to explain and contextualize its technical properties to their audience. As \molly put it:\begin{quote}
\textit{If you're making work that looks like sculpture, then who's your audience? Is it an audience who understands sculpture, but not what an algorithm is? Because I get real tired of explaining what an algorithm is.}
\end{quote}
Similarly, several artists initially resisted using computation because they did not share the values of existing technical communities, despite finding later success. 
\kim described a ``disconnect'' between the decision making processes of designers and developers. When \shantell worked in engineering communities, she struggled to reconcile her desires for transparent and aesthetically varied works with engineering norms that emphasized efficiency through technological opacity and uniformity. 
\lynn delayed pursuing coding because she was ``burned by the tech culture'' of a major Silicon Valley tech company when she worked there as a designer. Likewise, \molly described how she initially felt pushed to exhibit mastery in computer programming because of the ``power dynamics'' that exist between programming and domains like knitting or drawing:
\begin{quote}\textit{
There's a certain part of the population that's going think you're way cooler if you can code that thing than if you can draw it out, which is crazy.} 
\end{quote}
It is worthwhile to point out that the artists who experienced these conflicts were all women. The challenges they experienced are consistent with larger patterns of marginalization of certain groups---including women---in computer science and engineering~\cite{Margolis_Fisher_2002}.  Despite these conflicts, \kim, \shantell, \lynn and \molly persisted---often flourished---in computational production because they were able to find computational communities or engineering collaborators with similar values who prioritized what they had to say. \molly's immersion in traditional art communities strengthened her belief in the importance of manual and craft skills. She reached a point where she no longer felt like she needed to ``prove'' she could code well for her ``art to matter.''

\section{Discussion}
Here we discuss how artists currently engage with software in relationship to the current state of end-user programming (EUP) and creativity support tool (CST) research. From our analysis, we present critiques of existing approaches, design implications in response to these approaches, and strategies for new research opportunities in four categories. (1) Current tools that seek to lower the barrier of entry for creating art may rely on forms of automation and abstraction that hinder how artists traditionally learn through manual engagement with materials.
(2) Artists have complex and non-linear workflows that span physical and digital media; research can work towards unifying representations across both physical and digital objects.
(3) Artists are uniquely suited as technical collaborators in defining domain-specific programming representations and their contributions can expand the dimensions of what counts as systems research. 
(4) If systems engineering researchers wish to engage with artists, they have to consider not only the tools they build, but also the communities that surround them. \sonia{does them refer to tools / artists? something about the last point doesn't read very clearly.} 

\subsection{Automation Obscures Processes, Abstraction Obscures Data}
\changes{Based on our interview findings, we challenge 
the notion that computational automation reduces tedious manual labor. In section \ref{efficiency}, we described how forms of automation and abstraction that forgo manual control presented barriers to artists becoming self-reliant and producing aesthetically sophisticated outcomes. Artists instead relied on skilled manual execution and custom software to be efficient while preserving manual style. We argue that CST research can focus on not only making tasks faster or easier to accomplish, but also helping artists develop self-sufficiency through preserving manual control.}

While simpler controls make tasks more accessible to novices, who are the second most targeted user group in CST research \cite{frich-csts-10.1145/3290605.3300619}, forms of black-box automation can prevent artists from both using these tools in their existing workflows and in flexibly extending them across multiple workflows. For example, \molly had to write her own vinyl cutting algorithm from scratch because she could not edit the existing one provided with the software. As these forms of automation do not provide access to transparent algorithms, artists cannot adapt them in unique ways to develop idiosyncratic approaches to working. They instead may fall into aesthetics pre-determined by the tools, as described in section \ref{software}.

Furthermore, higher-level abstractions sometimes prevented artists from working at multiple granularities. \changes{For instance, \mac deliberately distinguished between the low-level representations of B\'ezier curves versus bitmaps when starting digital work as B\'ezier curves better afforded editing. In contrast, a higher-level representation, such as an automated effect or filter, would restrict this kind of meaningful decision making.} 
By depending on computational scaffolds that do not allow them to manually manipulate data representations, artists may not develop the skills to produce sophisticated artifacts. For instance, tools that use generative adversarial networks (GANs) such as ArtBreeder \cite{artbreeder} or various projects focusing on style transfer \cite{Semmo} forego manual control as artists can only specify input images; the engineers of such systems are the ones who determine the visual aesthetics of the final artifact. 

Inspired by the processes artists described in section \ref{beyond-artifacts}, forms of automation that do not remove manual control over processes can be applied to areas like exploration \cite{Hartmann_Yu_Allison_Yang_Klemmer_2008}, project management, and new ways of ``seeing'' and reflecting upon artwork \cite{fraser-livestream-10.1145/3313831.3376437}. CST research has also investigated data abstractions artists are familiar with, such as better ways of selecting layers \cite{layers-10.20380/GI2019.16} and undoing/redoing actions \cite{undo-10.1145/2856400.2856417, Myers-undo}. By allowing artists to have control over forms of automation and manipulate data representations, CSTs can focus on not only helping artists accomplish tasks, but also developing forms of self-reliance for unique outcomes.


\subsection{Adapting Digital Tools to the Unique Workflows of Artists}
In section \ref{physdigworkflows}, we described how artists have complex and non-linear workflows that move across a variety of mediums, both physical and digital. \changes{We draw from these workflows to advocate for programming systems that integrate manual input and physical materials with digital output and computational control. We also see design opportunities that take into account the aesthetic experience of using digital tools that capture the user's tacit knowledge.} \jingyi{is the emphasis in this section the ``programming'' part of systems? if so we should talk more about how specific affordances of programming languages can aid in supporting nonlinear workflows}

In creating work across physical and digital mediums, artists particularly highlighted the struggles of moving between separate tools and adapting their pieces to fit the constraints of the medium. In these transitions, artists experience what Winograd and Flores, in interpreting Heidegger, call ``breakdowns'' \cite{winograd1986understanding}---instances when artists engage with low-level properties of objects because they fail to accommodate fluid and invisible interactions. Our interviews showed examples of productive breakdowns when working with physical materials, such as \michael exploring many fabrication techniques and materials to transform his vector drawings, which ultimately inspired reflections to shape the final artwork. However, artists also experienced frustrating breakdowns that were simply a result of separate applications and tools not being able to interface with each other, such as \emily having to laboriously transfer her sketches across different digital mediums. 

\changes{We argue that one opportunity space for CST researchers---particularly in end-user programming---is to design program representations that support non-linearly moving between work spaces of physical and digital media, such that the output of any single stage can be the input of any other stage.} This is in contrast to current projects, such as those in digital fabrication \cite{forte-10.1145/3173574.3174070, savage-mm-10.1145/2807442.2807508, foldtronics-10.1145/3290605.3300858}, that assume more linear workflows: artists might start with a digital design tool, then use an existing program to compile it into machine code, and then fabricate it. \changes{For example, representations that integrate digital graphics, manual drawing gesture data, and CNC toolpaths could enable artists to program custom workflows across physical and digital forms of creation, and software environments that unify interfaces for authoring graphics, modifying manual input parameters, and programming CNC machine behavior could support rapid digital-physical transitions. Projects that use interactive paper substrates \cite{Garcia2012, bricosketch-10.1145/2817721.2817729} support domain experts in flexibly transitioning between the digital and physical.} Devendorf and Rosner state that ``hybridity,'' a melding of exactly two dichotomous categories, narrows the scope of what designers work with and privileges some interactions over others \cite{Devendorf_Rosner_2017}---we imagine designing data representations so artists can smoothly transition between all forms of making during any stage of their process.

Work has already investigated how to share workflows between different fabrication machines through integrated environments \cite{peek2018mods} or domain-specific languages \cite{tran-mom-10.1145/3332167.3356897}. Additionally, efforts in end-user programming \changes{like Enact~\cite{enact}} or Webstrates \cite{Klokmose_Eagan_Baader_Mackay_Beaudouin-Lafon_2015} have made progress on accommodating workflows that span multiple application programs. Webstrates aims to unify software representations at the level of the operating system such that objects can be shared across different applications that traditionally have their own internal representations. However, many forms of art production rely on specific affordances of physical materials that cannot be digitally replicated \cite{Ingold_2010}. In extending this concept to support the work of artists, we argue the ``operating system'' now has to include the physical spaces they also inhabit: for instance, from clay to 3D models to CAD software to G-Code to a 3D printer. Research efforts like Dynamicland \cite{dynamicland} have investigated these concepts by building a programming representation and operating system to unify actions in software, and those over space and time in the physical world. 


When working with physical materials, artists also based their tool choices not only on how they helped them execute visual artifacts but also on how they integrated into their varied intentions while making art. The same tool could be used by different artists for sketching, for refinement, or simply because it brought both tactile and emotional pleasure. The focus artists had on their feelings and senses when using tools suggests a space for systems researchers to pay attention to the emotional and aesthetic experiences of their software, in addition to targeting contributions that make accomplishing tasks faster or easier.

Finally, how artists interacted with tools was shaped by individual and embodied forms tacit knowledge, like knowing how much pressure to apply to a pen stroke while sketching versus lining. Capturing, formalizing, and evaluating these kinds of experiences---as well as emotional and aesthetic ones---is a challenge. Some CST researchers with backgrounds in art practice may draw from their own experiences \cite{torres10.1145/3325480.3325498}, but when researchers lack familiarity with the practices they want to support, artists can be powerful technical collaborators. 


\subsection{Visual Artists as Technical Collaborators}
As detailed in section \ref{software}, artists have a deep domain knowledge of their medium and appropriately choose to use or not use tools based on how the constraints and representations of the tool fit into their workflows. Devendorf et al. encountered similar knowledge in their residency with weavers, and argue that craftspeople should be considered technical collaborators with HCI researchers---while craftspeople's knowledge may be through a different medium than software engineering, that knowledge is still compatible with researchers' practices of design iteration and innovation \cite{Devendorf_Arquilla_Wirtanen_Anderson_Frost_2020}. We expand this notion of craftspeople as technical collaborators to visual artists; specifically, we argue visual artists are particularly well situated in \textit{defining domain-specific programming representations} that integrate manual expression with computational automation and digital manipulation. 

Artists who code understand the constraints of writing software and applying code towards artifacts they have previously manually created, such as \lynn writing Python scripts to speed up typographic labor or \chris deriving insights about which tools to build immediately after he manually finished his pieces. Because certain representations may be better aligned with painting, rather than drawing, or felting, or may even misalign with working by hand, we suggest that the domain expertise of artists, who are well versed in their craft, can help define such representations. This is in line with ideas from participatory and co-design \cite{muller1993participatory} methodologies; artists can make strong contributions in designing tools beyond the initial need-finding and final evaluation stages. \changes{However, beyond participatory design methods where artists help define high level features, we suggest that they also be involved in lower-level engineering discussions about data representations and implementation.} Artists are specifically interested in forging new and different paths because they open up new avenues for artistic exploration, instead of basing their success on efficiency. 


Research collaborations are a two-way street that should benefit all parties in tangible ways. Before starting collaborations, researchers should consider how HCI can also bring value to the types of achievements valued in an art career. For instance, artists and researchers have different incentives in disseminating tools they make---artists value releasing tools to their communities, while less than a quarter of CSTs are publicly available \cite{frich-csts-10.1145/3290605.3300619}, potentially due to the high value placed on the academic paper. We argue that HCI should broaden the scope of what is considered a systems engineering contribution to include the forms of output and inquiry artists value---such as making polished artifacts with systems over time, releasing useful technologies to support creative communities as opposed to constantly prioritizing novel systems, and investigating the art practices of individuals versus reporting on generalizable trends among large groups of artists. 

\subsection{Building Tool Communities}


In section \ref{motivations}, we described how artists were motivated to learn software development for various forms of power: in creating new functionality, intellectual growth, and technical legitimacy. In section \ref{aesthetics}, we described the experiences of four artists who overcame cultural barriers, differences in values, and feelings of exclusion to incorporate computation in their practices. Throughout this paper, we have argued for the value artists bring to our systems engineering research community. At the same time, these findings, as well as past research \cite{roque2016family}, show that the ways artists incorporate software and programming in their work is heavily influenced by the perceptions and values of their communities. If systems engineering researchers, as another technical community, seek to engage artists more broadly in developing tools, we also need to consider the communities surrounding the use of such tools.

For inspiration, we can look to two artist-led organizations that have been successful in teaching programming to artists: p5.js, a JavaScript port of the Processing creative coding language, and the School for Poetic Computation (SFPC), an artist-run school with the motto ``more poetry, less demo.'' Both these organizations have been successful in devoting many resources towards building communities where artists feel a sense of belonging, in addition to providing the tools to make novel work.
However, these groups---as well as the collaborations we reported on---represent a very narrow space within the larger software community. To support more artists, we need to find ways to build other communities elsewhere, as well as broaden our existing ones. 
The power of art comes from existing in intimate conversation with other humans, so leaving out such considerations of communities stifles the impact system builders can have on supporting artists working with computers.

\section{Conclusion}
In this paper, we examined the intersections of software and visual art to inform end-user programming and creativity support tools research. Our recommendations for software for creative production come from our thematic analysis of 13 artist interviews. By validating artists as core technical contributors in systems research and highlighting the need for inclusive community building around computational tools, we recognize and legitimize the value of distinct approaches to software use. Because systems that use high-level automation to make artwork conflict with artists' desires for fine-grained manipulations of their artifacts and tools, we suggest using automation to instead support analysis, organization, and reflection. We argue that diverse, non-linear physical-digital workflows can inform building flexible data representations for artwork and digital fabrication. Finally, we believe artists can contribute to shaping these representations because of their distinct approaches to how they learn, use, and build software that are rooted in humanistic inquiry and art practice.  
Based on these findings, our vision is that software \textit{for} artists will be written in close collaboration \textit{with} artists. Through these collaborations, we see great potential
to extend software use in creative practice and to grow inclusive communities around software development. 
\begin{acks}
The authors extend a huge gratitude to all the artists---\chrisfull, \emilyfull, \eranfull, \shantellfull, \miwafull, \fishfull, \mollyfull, \michaelfull, \macfull, \kimfull,  \benfull, \ninafull, and \lynnfull---without whom this work would not be possible. The authors would also like to thank colleagues Eric Rawn, Will Crichton, Evan Strasnick, Daniela Rosner, Laura Devendorf, Kristin Dew, Enric Boix-Adserà, and Kai Thaler for their valuable insights and conversations about this work. 
\end{acks}

\bibliographystyle{ACM-Reference-Format}
\bibliography{primary}


\begin{thebibliography}{101}


\ifx \showCODEN    \undefined \def \showCODEN     #1{\unskip}     \fi
\ifx \showDOI      \undefined \def \showDOI       #1{#1}\fi
\ifx \showISBNx    \undefined \def \showISBNx     #1{\unskip}     \fi
\ifx \showISBNxiii \undefined \def \showISBNxiii  #1{\unskip}     \fi
\ifx \showISSN     \undefined \def \showISSN      #1{\unskip}     \fi
\ifx \showLCCN     \undefined \def \showLCCN      #1{\unskip}     \fi
\ifx \shownote     \undefined \def \shownote      #1{#1}          \fi
\ifx \showarticletitle \undefined \def \showarticletitle #1{#1}   \fi
\ifx \showURL      \undefined \def \showURL       {\relax}        \fi
\providecommand\bibfield[2]{#2}
\providecommand\bibinfo[2]{#2}
\providecommand\natexlab[1]{#1}
\providecommand\showeprint[2][]{arXiv:#2}

\bibitem[\protect\citeauthoryear{{Barbosa}, {Wanderley}, and {Huot}}{{Barbosa}
  et~al\mbox{.}}{2018}]%
        {Barbosa2018}
\bibfield{author}{\bibinfo{person}{J. {Barbosa}}, \bibinfo{person}{M.~M.
  {Wanderley}}, {and} \bibinfo{person}{S. {Huot}}.}
  \bibinfo{year}{2018}\natexlab{}.
\newblock \showarticletitle{ZenStates: Easy-to-Understand Yet Expressive
  Specifications for Creative Interactive Environments}. In
  \bibinfo{booktitle}{\emph{2018 IEEE Symposium on Visual Languages and
  Human-Centric Computing (VL/HCC)}}. \bibinfo{pages}{167--175}.
\newblock
\urldef\tempurl%
\url{https://doi.org/10.1109/VLHCC.2018.8506491}
\showDOI{\tempurl}


\bibitem[\protect\citeauthoryear{Bardzell, Rosner, and Bardzell}{Bardzell
  et~al\mbox{.}}{2012}]%
        {Bardzell_Rosner_Bardzell_2012}
\bibfield{author}{\bibinfo{person}{Shaowen Bardzell},
  \bibinfo{person}{Daniela~K. Rosner}, {and} \bibinfo{person}{Jeffrey
  Bardzell}.} \bibinfo{year}{2012}\natexlab{}.
\newblock \showarticletitle{Crafting quality in design: integrity, creativity,
  and public sensibility}. In \bibinfo{booktitle}{\emph{Proceedings of the
  Designing Interactive Systems Conference on - DIS ’12}}.
  \bibinfo{publisher}{ACM Press}, \bibinfo{pages}{11}.
\newblock
\showISBNx{978-1-4503-1210-3}
\urldef\tempurl%
\url{https://doi.org/10.1145/2317956.2317959}
\showDOI{\tempurl}


\bibitem[\protect\citeauthoryear{Barnes, Jacobs, Sanders, Goldman,
  Rusinkiewicz, Finkelstein, and Agrawala}{Barnes et~al\mbox{.}}{2008}]%
        {Barnes-video-puppetry}
\bibfield{author}{\bibinfo{person}{Connelly Barnes}, \bibinfo{person}{David~E.
  Jacobs}, \bibinfo{person}{Jason Sanders}, \bibinfo{person}{Dan~B Goldman},
  \bibinfo{person}{Szymon Rusinkiewicz}, \bibinfo{person}{Adam Finkelstein},
  {and} \bibinfo{person}{Maneesh Agrawala}.} \bibinfo{year}{2008}\natexlab{}.
\newblock \showarticletitle{Video Puppetry: A Performative Interface for Cutout
  Animation}.
\newblock \bibinfo{journal}{\emph{ACM Trans. Graph.}} \bibinfo{volume}{27},
  \bibinfo{number}{5}, Article \bibinfo{articleno}{124} (\bibinfo{date}{Dec.}
  \bibinfo{year}{2008}), \bibinfo{numpages}{9}~pages.
\newblock
\showISSN{0730-0301}
\urldef\tempurl%
\url{https://doi.org/10.1145/1409060.1409077}
\showDOI{\tempurl}


\bibitem[\protect\citeauthoryear{Batra, Phogat, and Beri}{Batra
  et~al\mbox{.}}{2019}]%
        {Batra-layout}
\bibfield{author}{\bibinfo{person}{Vineet Batra}, \bibinfo{person}{Ankit
  Phogat}, {and} \bibinfo{person}{Tarun Beri}.}
  \bibinfo{year}{2019}\natexlab{}.
\newblock \showarticletitle{Massively Parallel Layout Generation in Real Time}.
  In \bibinfo{booktitle}{\emph{ACM SIGGRAPH 2019 Posters}} (Los Angeles,
  California) \emph{(\bibinfo{series}{SIGGRAPH '19})}.
  \bibinfo{publisher}{Association for Computing Machinery},
  \bibinfo{address}{New York, NY, USA}, Article \bibinfo{articleno}{3},
  \bibinfo{numpages}{2}~pages.
\newblock
\showISBNx{9781450363143}
\urldef\tempurl%
\url{https://doi.org/10.1145/3306214.3338596}
\showDOI{\tempurl}


\bibitem[\protect\citeauthoryear{Beaudouin-Lafon}{Beaudouin-Lafon}{2017}]%
        {Beaudouin-Lafon_2017}
\bibfield{author}{\bibinfo{person}{Michel Beaudouin-Lafon}.}
  \bibinfo{year}{2017}\natexlab{}.
\newblock \showarticletitle{Towards Unified Principles of Interaction}. In
  \bibinfo{booktitle}{\emph{Proceedings of the 12th Biannual Conference on
  Italian SIGCHI Chapter}} \emph{(\bibinfo{series}{CHItaly ’17})}.
  \bibinfo{publisher}{Association for Computing Machinery},
  \bibinfo{pages}{1–2}.
\newblock
\showISBNx{978-1-4503-5237-6}
\urldef\tempurl%
\url{https://doi.org/10.1145/3125571.3125602}
\showDOI{\tempurl}


\bibitem[\protect\citeauthoryear{Berger}{Berger}{2014}]%
        {Berger}
\bibfield{author}{\bibinfo{person}{John Berger}.}
  \bibinfo{year}{2014}\natexlab{}.
\newblock \bibinfo{booktitle}{\emph{Selected Essays of John Berger}}.
\newblock \bibinfo{publisher}{Bloomsbury Publishing Plc}.
\newblock


\bibitem[\protect\citeauthoryear{Bernstein and Li}{Bernstein and Li}{2015}]%
        {Bernstein-lillicon}
\bibfield{author}{\bibinfo{person}{Gilbert~Louis Bernstein} {and}
  \bibinfo{person}{Wilmot Li}.} \bibinfo{year}{2015}\natexlab{}.
\newblock \showarticletitle{Lillicon: Using Transient Widgets to Create Scale
  Variations of Icons}.
\newblock \bibinfo{journal}{\emph{ACM Trans. Graph.}} \bibinfo{volume}{34},
  \bibinfo{number}{4}, Article \bibinfo{articleno}{144} (\bibinfo{date}{July}
  \bibinfo{year}{2015}), \bibinfo{numpages}{11}~pages.
\newblock
\showISSN{0730-0301}
\urldef\tempurl%
\url{https://doi.org/10.1145/2766980}
\showDOI{\tempurl}


\bibitem[\protect\citeauthoryear{Brandt, Guo, Lewenstein, and Klemmer}{Brandt
  et~al\mbox{.}}{2008}]%
        {Brandt_Guo_Lewenstein_Klemmer_2008}
\bibfield{author}{\bibinfo{person}{Joel Brandt}, \bibinfo{person}{Philip~J.
  Guo}, \bibinfo{person}{Joel Lewenstein}, {and} \bibinfo{person}{Scott~R.
  Klemmer}.} \bibinfo{year}{2008}\natexlab{}.
\newblock \showarticletitle{Opportunistic programming: how rapid ideation and
  prototyping occur in practice}. In \bibinfo{booktitle}{\emph{Proceedings of
  the 4th international workshop on End-user software engineering - WEUSE
  ’08}}. \bibinfo{publisher}{ACM Press}, \bibinfo{pages}{1–5}.
\newblock
\showISBNx{978-1-60558-034-0}
\urldef\tempurl%
\url{https://doi.org/10.1145/1370847.1370848}
\showDOI{\tempurl}


\bibitem[\protect\citeauthoryear{Braun and Clarke}{Braun and Clarke}{2006}]%
        {braun}
\bibfield{author}{\bibinfo{person}{Virginia Braun} {and}
  \bibinfo{person}{Victoria Clarke}.} \bibinfo{year}{2006}\natexlab{}.
\newblock \showarticletitle{Using thematic analysis in psychology}.
\newblock \bibinfo{journal}{\emph{Qualitative Research in Psychology}}
  \bibinfo{volume}{3}, \bibinfo{number}{2} (\bibinfo{year}{2006}),
  \bibinfo{pages}{77--101}.
\newblock
\urldef\tempurl%
\url{https://doi.org/10.1191/1478088706qp063oa}
\showDOI{\tempurl}
\showeprint{https://www.tandfonline.com/doi/pdf/10.1191/1478088706qp063oa}


\bibitem[\protect\citeauthoryear{Braun and Clarke}{Braun and Clarke}{2019}]%
        {braun_reflexive}
\bibfield{author}{\bibinfo{person}{Virginia Braun} {and}
  \bibinfo{person}{Victoria Clarke}.} \bibinfo{year}{2019}\natexlab{}.
\newblock \showarticletitle{Reflecting on reflexive thematic analysis}.
\newblock \bibinfo{journal}{\emph{Qualitative Research in Sport, Exercise and
  Health}} \bibinfo{volume}{11}, \bibinfo{number}{4} (\bibinfo{year}{2019}),
  \bibinfo{pages}{589--597}.
\newblock
\urldef\tempurl%
\url{https://doi.org/10.1080/2159676X.2019.1628806}
\showDOI{\tempurl}
\showeprint{https://doi.org/10.1080/2159676X.2019.1628806}


\bibitem[\protect\citeauthoryear{Champandard}{Champandard}{2016}]%
        {Champandard}
\bibfield{author}{\bibinfo{person}{Alex~J. Champandard}.}
  \bibinfo{year}{2016}\natexlab{}.
\newblock \showarticletitle{Semantic Style Transfer and Turning Two-Bit Doodles
  into Fine Artworks}.
\newblock \bibinfo{journal}{\emph{CoRR}}  \bibinfo{volume}{abs/1603.01768}
  (\bibinfo{year}{2016}).
\newblock
\showeprint[arxiv]{1603.01768}
\urldef\tempurl%
\url{http://arxiv.org/abs/1603.01768}
\showURL{%
\tempurl}


\bibitem[\protect\citeauthoryear{Cheatle and Jackson}{Cheatle and
  Jackson}{2015}]%
        {Cheatle_Jackson_2015}
\bibfield{author}{\bibinfo{person}{Amy Cheatle} {and}
  \bibinfo{person}{Steven~J. Jackson}.} \bibinfo{year}{2015}\natexlab{}.
\newblock \showarticletitle{Digital Entanglements: Craft, Computation and
  Collaboration in Fine Art Furniture Production}. In
  \bibinfo{booktitle}{\emph{Proceedings of the 18th ACM Conference on Computer
  Supported Cooperative Work \& Social Computing}} \emph{(\bibinfo{series}{CSCW
  ’15})}. \bibinfo{publisher}{Association for Computing Machinery},
  \bibinfo{pages}{958–968}.
\newblock
\showISBNx{978-1-4503-2922-4}
\urldef\tempurl%
\url{https://doi.org/10.1145/2675133.2675291}
\showDOI{\tempurl}


\bibitem[\protect\citeauthoryear{Chen, Wei, Hartmann, and Agrawala}{Chen
  et~al\mbox{.}}{2016}]%
        {undo-10.1145/2856400.2856417}
\bibfield{author}{\bibinfo{person}{Hsiang-Ting Chen}, \bibinfo{person}{Li-Yi
  Wei}, \bibinfo{person}{Bj\"{o}rn Hartmann}, {and} \bibinfo{person}{Maneesh
  Agrawala}.} \bibinfo{year}{2016}\natexlab{}.
\newblock \showarticletitle{Data-Driven Adaptive History for Image Editing}. In
  \bibinfo{booktitle}{\emph{Proceedings of the 20th ACM SIGGRAPH Symposium on
  Interactive 3D Graphics and Games}} (Redmond, Washington)
  \emph{(\bibinfo{series}{I3D '16})}. \bibinfo{publisher}{Association for
  Computing Machinery}, \bibinfo{address}{New York, NY, USA},
  \bibinfo{pages}{103–111}.
\newblock
\showISBNx{9781450340434}
\urldef\tempurl%
\url{https://doi.org/10.1145/2856400.2856417}
\showDOI{\tempurl}


\bibitem[\protect\citeauthoryear{Chen, Tao, Wang, Kang, Grossman, Coros, and
  Hudson}{Chen et~al\mbox{.}}{2018}]%
        {forte-10.1145/3173574.3174070}
\bibfield{author}{\bibinfo{person}{Xiang~'Anthony' Chen}, \bibinfo{person}{Ye
  Tao}, \bibinfo{person}{Guanyun Wang}, \bibinfo{person}{Runchang Kang},
  \bibinfo{person}{Tovi Grossman}, \bibinfo{person}{Stelian Coros}, {and}
  \bibinfo{person}{Scott~E. Hudson}.} \bibinfo{year}{2018}\natexlab{}.
\newblock \showarticletitle{Forte: User-Driven Generative Design}. In
  \bibinfo{booktitle}{\emph{Proceedings of the 2018 CHI Conference on Human
  Factors in Computing Systems}} (Montreal QC, Canada)
  \emph{(\bibinfo{series}{CHI '18})}. \bibinfo{publisher}{Association for
  Computing Machinery}, \bibinfo{address}{New York, NY, USA},
  \bibinfo{pages}{1–12}.
\newblock
\showISBNx{9781450356206}
\urldef\tempurl%
\url{https://doi.org/10.1145/3173574.3174070}
\showDOI{\tempurl}


\bibitem[\protect\citeauthoryear{Chugh, Hempel, Spradlin, and Albers}{Chugh
  et~al\mbox{.}}{2016}]%
        {Chugh-2980983.2908103}
\bibfield{author}{\bibinfo{person}{Ravi Chugh}, \bibinfo{person}{Brian Hempel},
  \bibinfo{person}{Mitchell Spradlin}, {and} \bibinfo{person}{Jacob Albers}.}
  \bibinfo{year}{2016}\natexlab{}.
\newblock \showarticletitle{Programmatic and Direct Manipulation, Together at
  Last}.
\newblock \bibinfo{journal}{\emph{SIGPLAN Not.}} \bibinfo{volume}{51},
  \bibinfo{number}{6} (\bibinfo{date}{June} \bibinfo{year}{2016}),
  \bibinfo{pages}{341–354}.
\newblock
\showISSN{0362-1340}
\urldef\tempurl%
\url{https://doi.org/10.1145/2980983.2908103}
\showDOI{\tempurl}


\bibitem[\protect\citeauthoryear{Cycling74}{Cycling74}{2017}]%
        {max}
\bibfield{author}{\bibinfo{person}{Cycling74}.}
  \bibinfo{year}{2017}\natexlab{}.
\newblock \bibinfo{booktitle}{\emph{Max}}.
\newblock
\newblock
\shownote{\url{https://cycling74.com/products/max/}.}


\bibitem[\protect\citeauthoryear{Devendorf, Arquilla, Wirtanen, Anderson, and
  Frost}{Devendorf et~al\mbox{.}}{2020}]%
        {Devendorf_Arquilla_Wirtanen_Anderson_Frost_2020}
\bibfield{author}{\bibinfo{person}{Laura Devendorf}, \bibinfo{person}{Katya
  Arquilla}, \bibinfo{person}{Sandra Wirtanen}, \bibinfo{person}{Allison
  Anderson}, {and} \bibinfo{person}{Steven Frost}.}
  \bibinfo{year}{2020}\natexlab{}.
\newblock \showarticletitle{Craftspeople as Technical Collaborators: Lessons
  Learned through an Experimental Weaving Residency}. In
  \bibinfo{booktitle}{\emph{Proceedings of the 2020 CHI Conference on Human
  Factors in Computing Systems}} \emph{(\bibinfo{series}{CHI ’20})}.
  \bibinfo{publisher}{Association for Computing Machinery},
  \bibinfo{pages}{1–13}.
\newblock
\showISBNx{978-1-4503-6708-0}
\urldef\tempurl%
\url{https://doi.org/10.1145/3313831.3376820}
\showDOI{\tempurl}


\bibitem[\protect\citeauthoryear{Devendorf and Rosner}{Devendorf and
  Rosner}{2017}]%
        {Devendorf_Rosner_2017}
\bibfield{author}{\bibinfo{person}{Laura Devendorf} {and}
  \bibinfo{person}{Daniela~K. Rosner}.} \bibinfo{year}{2017}\natexlab{}.
\newblock \showarticletitle{Beyond Hybrids: Metaphors and Margins in Design}.
  In \bibinfo{booktitle}{\emph{Proceedings of the 2017 Conference on Designing
  Interactive Systems - DIS ’17}}. \bibinfo{publisher}{ACM Press},
  \bibinfo{pages}{995–1000}.
\newblock
\showISBNx{978-1-4503-4922-2}
\urldef\tempurl%
\url{https://doi.org/10.1145/3064663.3064705}
\showDOI{\tempurl}


\bibitem[\protect\citeauthoryear{Devendorf and Ryokai}{Devendorf and
  Ryokai}{2015}]%
        {beingmachine-10.1145/2702123.2702547}
\bibfield{author}{\bibinfo{person}{Laura Devendorf} {and}
  \bibinfo{person}{Kimiko Ryokai}.} \bibinfo{year}{2015}\natexlab{}.
\newblock \showarticletitle{Being the Machine: Reconfiguring Agency and Control
  in Hybrid Fabrication}. In \bibinfo{booktitle}{\emph{Proceedings of the 33rd
  Annual ACM Conference on Human Factors in Computing Systems}} (Seoul,
  Republic of Korea) \emph{(\bibinfo{series}{CHI '15})}.
  \bibinfo{publisher}{Association for Computing Machinery},
  \bibinfo{address}{New York, NY, USA}, \bibinfo{pages}{2477–2486}.
\newblock
\showISBNx{9781450331456}
\urldef\tempurl%
\url{https://doi.org/10.1145/2702123.2702547}
\showDOI{\tempurl}


\bibitem[\protect\citeauthoryear{Dew and Rosner}{Dew and Rosner}{2018}]%
        {Dew_Rosner_2018}
\bibfield{author}{\bibinfo{person}{Kristin~N. Dew} {and}
  \bibinfo{person}{Daniela~K. Rosner}.} \bibinfo{year}{2018}\natexlab{}.
\newblock \showarticletitle{Lessons from the Woodshop: Cultivating Design with
  Living Materials}. In \bibinfo{booktitle}{\emph{Proceedings of the 2018 CHI
  Conference on Human Factors in Computing Systems - CHI ’18}}.
  \bibinfo{publisher}{ACM Press}, \bibinfo{pages}{1–12}.
\newblock
\showISBNx{978-1-4503-5620-6}
\urldef\tempurl%
\url{https://doi.org/10.1145/3173574.3174159}
\showDOI{\tempurl}


\bibitem[\protect\citeauthoryear{Do and Gross}{Do and Gross}{2007}]%
        {Do_Gross}
\bibfield{author}{\bibinfo{person}{Ellen Yi-Luen Do} {and}
  \bibinfo{person}{Mark~D. Gross}.} \bibinfo{year}{2007}\natexlab{}.
\newblock \showarticletitle{Environments for Creativity: A Lab for Making
  Things}. In \bibinfo{booktitle}{\emph{Proceedings of the 6th ACM SIGCHI
  Conference on Creativity \& Cognition}} (Washington, DC, USA)
  \emph{(\bibinfo{series}{C\&C '07})}. \bibinfo{publisher}{Association for
  Computing Machinery}, \bibinfo{address}{New York, NY, USA},
  \bibinfo{pages}{27–36}.
\newblock
\showISBNx{9781595937124}
\urldef\tempurl%
\url{https://doi.org/10.1145/1254960.1254965}
\showDOI{\tempurl}


\bibitem[\protect\citeauthoryear{E, Fried, Lu, Zhang, Mech, Echevarria,
  Hanrahan, and Landay}{E et~al\mbox{.}}{2020}]%
        {E-photography}
\bibfield{author}{\bibinfo{person}{Jane~L. E}, \bibinfo{person}{Ohad Fried},
  \bibinfo{person}{Jingwan Lu}, \bibinfo{person}{Jianming Zhang},
  \bibinfo{person}{Radom\'{\i}r Mech}, \bibinfo{person}{Jose Echevarria},
  \bibinfo{person}{Pat Hanrahan}, {and} \bibinfo{person}{James~A. Landay}.}
  \bibinfo{year}{2020}\natexlab{}.
\newblock \showarticletitle{Adaptive Photographic Composition Guidance}. In
  \bibinfo{booktitle}{\emph{Proceedings of the 2020 CHI Conference on Human
  Factors in Computing Systems}} (Honolulu, HI, USA)
  \emph{(\bibinfo{series}{CHI '20})}. \bibinfo{publisher}{Association for
  Computing Machinery}, \bibinfo{address}{New York, NY, USA},
  \bibinfo{pages}{1–13}.
\newblock
\showISBNx{9781450367080}
\urldef\tempurl%
\url{https://doi.org/10.1145/3313831.3376635}
\showDOI{\tempurl}


\bibitem[\protect\citeauthoryear{Eaglestone, Ford, Nuhn, Moore, and
  Brown}{Eaglestone et~al\mbox{.}}{2001}]%
        {Eaglestone2001}
\bibfield{author}{\bibinfo{person}{Barry Eaglestone}, \bibinfo{person}{Nigel
  Ford}, \bibinfo{person}{Ralf Nuhn}, \bibinfo{person}{Adrian Moore}, {and}
  \bibinfo{person}{Guy~J Brown}.} \bibinfo{year}{2001}\natexlab{}.
\newblock \showarticletitle{Composition systems requirements for creativity:
  what research methodology}. In \bibinfo{booktitle}{\emph{In Proc. MOSART
  Workshop}}. \bibinfo{pages}{7--16}.
\newblock


\bibitem[\protect\citeauthoryear{Feist}{Feist}{1999}]%
        {feist1999}
\bibfield{author}{\bibinfo{person}{Gregory~J Feist}.}
  \bibinfo{year}{1999}\natexlab{}.
\newblock \showarticletitle{The influence of personality on artistic and
  scientific creativity}.
\newblock \bibinfo{journal}{\emph{Handbook of creativity}}
  (\bibinfo{year}{1999}), \bibinfo{pages}{273}.
\newblock


\bibitem[\protect\citeauthoryear{Fiebrink and Sonami}{Fiebrink and
  Sonami}{[n.d.]}]%
        {Fiebrink_Sonami}
\bibfield{author}{\bibinfo{person}{Rebecca Fiebrink} {and}
  \bibinfo{person}{Laetitia Sonami}.} \bibinfo{year}{[n.d.]}\natexlab{}.
\newblock \showarticletitle{Reflections on Eight Years of Instrument Creation
  with Machine Learning}.
\newblock  (\bibinfo{year}{[n.\,d.]}), \bibinfo{pages}{6}.
\newblock


\bibitem[\protect\citeauthoryear{Fraser, Kim, Shin, Brandt, and
  Dontcheva}{Fraser et~al\mbox{.}}{2020}]%
        {fraser-livestream-10.1145/3313831.3376437}
\bibfield{author}{\bibinfo{person}{C.~Ailie Fraser}, \bibinfo{person}{Joy~O.
  Kim}, \bibinfo{person}{Hijung~Valentina Shin}, \bibinfo{person}{Joel Brandt},
  {and} \bibinfo{person}{Mira Dontcheva}.} \bibinfo{year}{2020}\natexlab{}.
\newblock \showarticletitle{Temporal Segmentation of Creative Live Streams}. In
  \bibinfo{booktitle}{\emph{Proceedings of the 2020 CHI Conference on Human
  Factors in Computing Systems}} (Honolulu, HI, USA)
  \emph{(\bibinfo{series}{CHI '20})}. \bibinfo{publisher}{Association for
  Computing Machinery}, \bibinfo{address}{New York, NY, USA},
  \bibinfo{pages}{1–12}.
\newblock
\showISBNx{9781450367080}
\urldef\tempurl%
\url{https://doi.org/10.1145/3313831.3376437}
\showDOI{\tempurl}


\bibitem[\protect\citeauthoryear{Frich, MacDonald~Vermeulen, Remy, Biskjaer,
  and Dalsgaard}{Frich et~al\mbox{.}}{2019}]%
        {frich-csts-10.1145/3290605.3300619}
\bibfield{author}{\bibinfo{person}{Jonas Frich}, \bibinfo{person}{Lindsay
  MacDonald~Vermeulen}, \bibinfo{person}{Christian Remy},
  \bibinfo{person}{Michael~Mose Biskjaer}, {and} \bibinfo{person}{Peter
  Dalsgaard}.} \bibinfo{year}{2019}\natexlab{}.
\newblock \showarticletitle{Mapping the Landscape of Creativity Support Tools
  in HCI}. In \bibinfo{booktitle}{\emph{Proceedings of the 2019 CHI Conference
  on Human Factors in Computing Systems}} (Glasgow, Scotland Uk)
  \emph{(\bibinfo{series}{CHI '19})}. \bibinfo{publisher}{Association for
  Computing Machinery}, \bibinfo{address}{New York, NY, USA},
  \bibinfo{pages}{1–18}.
\newblock
\showISBNx{9781450359702}
\urldef\tempurl%
\url{https://doi.org/10.1145/3290605.3300619}
\showDOI{\tempurl}


\bibitem[\protect\citeauthoryear{Fujii}{Fujii}{2017}]%
        {fujii2017interviewing}
\bibfield{author}{\bibinfo{person}{Lee~Ann Fujii}.}
  \bibinfo{year}{2017}\natexlab{}.
\newblock \bibinfo{booktitle}{\emph{Interviewing in social science research: A
  relational approach}}.
\newblock \bibinfo{publisher}{Routledge}.
\newblock


\bibitem[\protect\citeauthoryear{Garcia, Tsandilas, Agon, and Mackay}{Garcia
  et~al\mbox{.}}{2012}]%
        {Garcia2012}
\bibfield{author}{\bibinfo{person}{J\'{e}r\'{e}mie Garcia},
  \bibinfo{person}{Theophanis Tsandilas}, \bibinfo{person}{Carlos Agon}, {and}
  \bibinfo{person}{Wendy Mackay}.} \bibinfo{year}{2012}\natexlab{}.
\newblock \showarticletitle{Interactive Paper Substrates to Support Musical
  Creation}. In \bibinfo{booktitle}{\emph{Proceedings of the SIGCHI Conference
  on Human Factors in Computing Systems}} (Austin, Texas, USA)
  \emph{(\bibinfo{series}{CHI '12})}. \bibinfo{publisher}{Association for
  Computing Machinery}, \bibinfo{address}{New York, NY, USA},
  \bibinfo{pages}{1825–1828}.
\newblock
\showISBNx{9781450310154}
\urldef\tempurl%
\url{https://doi.org/10.1145/2207676.2208316}
\showDOI{\tempurl}


\bibitem[\protect\citeauthoryear{Gatys, Ecker, and Bethge}{Gatys
  et~al\mbox{.}}{2015}]%
        {Gatys}
\bibfield{author}{\bibinfo{person}{Leon~A. Gatys},
  \bibinfo{person}{Alexander~S. Ecker}, {and} \bibinfo{person}{M. Bethge}.}
  \bibinfo{year}{2015}\natexlab{}.
\newblock \showarticletitle{A Neural Algorithm of Artistic Style}.
\newblock \bibinfo{journal}{\emph{ArXiv}}  \bibinfo{volume}{abs/1508.06576}
  (\bibinfo{year}{2015}).
\newblock


\bibitem[\protect\citeauthoryear{Goodman}{Goodman}{1968}]%
        {goodman1968languages}
\bibfield{author}{\bibinfo{person}{Nelson Goodman}.}
  \bibinfo{year}{1968}\natexlab{}.
\newblock \showarticletitle{Languages of Art: An Approach to a Theory of
  Symbols. The Bobbs-Merrill Company}.
\newblock \bibinfo{journal}{\emph{Inc. New York, Indianapolis}}
  (\bibinfo{year}{1968}).
\newblock


\bibitem[\protect\citeauthoryear{Gross}{Gross}{2009}]%
        {Gross}
\bibfield{author}{\bibinfo{person}{M.~D. Gross}.}
  \bibinfo{year}{2009}\natexlab{}.
\newblock \showarticletitle{Visual Languages and Visual Thinking: Sketch Based
  Interaction and Modeling}. In \bibinfo{booktitle}{\emph{Proceedings of the
  6th Eurographics Symposium on Sketch-Based Interfaces and Modeling}} (New
  Orleans, Louisiana) \emph{(\bibinfo{series}{SBIM '09})}.
  \bibinfo{publisher}{Association for Computing Machinery},
  \bibinfo{address}{New York, NY, USA}, \bibinfo{pages}{7–11}.
\newblock
\showISBNx{9781605586021}
\urldef\tempurl%
\url{https://doi.org/10.1145/1572741.1572743}
\showDOI{\tempurl}


\bibitem[\protect\citeauthoryear{Hartmann, Yu, Allison, Yang, and
  Klemmer}{Hartmann et~al\mbox{.}}{2008}]%
        {Hartmann_Yu_Allison_Yang_Klemmer_2008}
\bibfield{author}{\bibinfo{person}{Björn Hartmann}, \bibinfo{person}{Loren
  Yu}, \bibinfo{person}{Abel Allison}, \bibinfo{person}{Yeonsoo Yang}, {and}
  \bibinfo{person}{Scott~R. Klemmer}.} \bibinfo{year}{2008}\natexlab{}.
\newblock \showarticletitle{Design as exploration: creating interface
  alternatives through parallel authoring and runtime tuning}. In
  \bibinfo{booktitle}{\emph{Proceedings of the 21st annual ACM symposium on
  User interface software and technology - UIST ’08}}.
  \bibinfo{publisher}{ACM Press}, \bibinfo{pages}{91}.
\newblock
\showISBNx{978-1-59593-975-3}
\urldef\tempurl%
\url{https://doi.org/10.1145/1449715.1449732}
\showDOI{\tempurl}


\bibitem[\protect\citeauthoryear{Hempel, Lubin, and Chugh}{Hempel
  et~al\mbox{.}}{2019}]%
        {Hempel-3332165.3347925}
\bibfield{author}{\bibinfo{person}{Brian Hempel}, \bibinfo{person}{Justin
  Lubin}, {and} \bibinfo{person}{Ravi Chugh}.} \bibinfo{year}{2019}\natexlab{}.
\newblock \showarticletitle{Sketch-n-Sketch: Output-Directed Programming for
  SVG}. In \bibinfo{booktitle}{\emph{Proceedings of the 32nd Annual ACM
  Symposium on User Interface Software and Technology}} (New Orleans, LA, USA)
  \emph{(\bibinfo{series}{UIST '19})}. \bibinfo{publisher}{Association for
  Computing Machinery}, \bibinfo{address}{New York, NY, USA},
  \bibinfo{pages}{281–292}.
\newblock
\showISBNx{9781450368162}
\urldef\tempurl%
\url{https://doi.org/10.1145/3332165.3347925}
\showDOI{\tempurl}


\bibitem[\protect\citeauthoryear{Hertzmann}{Hertzmann}{2018}]%
        {Hertzmann}
\bibfield{author}{\bibinfo{person}{Aaron Hertzmann}.}
  \bibinfo{year}{2018}\natexlab{}.
\newblock \showarticletitle{Can Computers Create Art?}
\newblock \bibinfo{journal}{\emph{CoRR}}  \bibinfo{volume}{abs/1801.04486}
  (\bibinfo{year}{2018}).
\newblock
\showeprint[arxiv]{1801.04486}
\urldef\tempurl%
\url{http://arxiv.org/abs/1801.04486}
\showURL{%
\tempurl}


\bibitem[\protect\citeauthoryear{Iizuka, Simo-Serra, and Ishikawa}{Iizuka
  et~al\mbox{.}}{2016}]%
        {Iizuka}
\bibfield{author}{\bibinfo{person}{Satoshi Iizuka}, \bibinfo{person}{Edgar
  Simo-Serra}, {and} \bibinfo{person}{Hiroshi Ishikawa}.}
  \bibinfo{year}{2016}\natexlab{}.
\newblock \showarticletitle{Let There Be Color! Joint End-to-End Learning of
  Global and Local Image Priors for Automatic Image Colorization with
  Simultaneous Classification}.
\newblock \bibinfo{journal}{\emph{ACM Trans. Graph.}} \bibinfo{volume}{35},
  \bibinfo{number}{4}, Article \bibinfo{articleno}{110} (\bibinfo{date}{July}
  \bibinfo{year}{2016}), \bibinfo{numpages}{11}~pages.
\newblock
\showISSN{0730-0301}
\urldef\tempurl%
\url{https://doi.org/10.1145/2897824.2925974}
\showDOI{\tempurl}


\bibitem[\protect\citeauthoryear{Ingold}{Ingold}{2010}]%
        {Ingold_2010}
\bibfield{author}{\bibinfo{person}{T. Ingold}.}
  \bibinfo{year}{2010}\natexlab{}.
\newblock \showarticletitle{The textility of making}.
\newblock \bibinfo{journal}{\emph{Cambridge Journal of Economics}}
  \bibinfo{volume}{34}, \bibinfo{number}{1} (\bibinfo{date}{Jan}
  \bibinfo{year}{2010}), \bibinfo{pages}{91–102}.
\newblock
\showISSN{0309-166X, 1464-3545}
\urldef\tempurl%
\url{https://doi.org/10.1093/cje/bep042}
\showDOI{\tempurl}


\bibitem[\protect\citeauthoryear{Jacobs}{Jacobs}{2018}]%
        {jacobs-sfpc}
\bibfield{author}{\bibinfo{person}{J. Jacobs}.}
  \bibinfo{year}{2018}\natexlab{}.
\newblock \bibinfo{booktitle}{\emph{SFPC Residency Reflections}}.
\newblock
\newblock
\shownote{\url{https://medium.com/sfpc/sfpc-residency-reflections-bf32204c92aa}.}


\bibitem[\protect\citeauthoryear{Jacobs, Brandt, Mech, and Resnick}{Jacobs
  et~al\mbox{.}}{2018}]%
        {Jacobs_Brandt_Mech_Resnick_2018}
\bibfield{author}{\bibinfo{person}{Jennifer Jacobs}, \bibinfo{person}{Joel
  Brandt}, \bibinfo{person}{Radomír Mech}, {and} \bibinfo{person}{Mitchel
  Resnick}.} \bibinfo{year}{2018}\natexlab{}.
\newblock \showarticletitle{Extending Manual Drawing Practices with
  Artist-Centric Programming Tools}. In \bibinfo{booktitle}{\emph{Proceedings
  of the 2018 CHI Conference on Human Factors in Computing Systems}}
  \emph{(\bibinfo{series}{CHI ’18})}. \bibinfo{publisher}{Association for
  Computing Machinery}, \bibinfo{pages}{1–13}.
\newblock
\showISBNx{978-1-4503-5620-6}
\urldef\tempurl%
\url{https://doi.org/10.1145/3173574.3174164}
\showDOI{\tempurl}


\bibitem[\protect\citeauthoryear{Jacobs, Gogia, Mundefinedch, and
  Brandt}{Jacobs et~al\mbox{.}}{2017}]%
        {Jacobs-para}
\bibfield{author}{\bibinfo{person}{Jennifer Jacobs}, \bibinfo{person}{Sumit
  Gogia}, \bibinfo{person}{Radom\'{\i}r Mundefinedch}, {and}
  \bibinfo{person}{Joel~R. Brandt}.} \bibinfo{year}{2017}\natexlab{}.
\newblock \showarticletitle{Supporting Expressive Procedural Art Creation
  through Direct Manipulation}. In \bibinfo{booktitle}{\emph{Proceedings of the
  2017 CHI Conference on Human Factors in Computing Systems}} (Denver,
  Colorado, USA) \emph{(\bibinfo{series}{CHI '17})}.
  \bibinfo{publisher}{Association for Computing Machinery},
  \bibinfo{address}{New York, NY, USA}, \bibinfo{pages}{6330–6341}.
\newblock
\showISBNx{9781450346559}
\urldef\tempurl%
\url{https://doi.org/10.1145/3025453.3025927}
\showDOI{\tempurl}


\bibitem[\protect\citeauthoryear{Jacobs and Zoran}{Jacobs and Zoran}{2015}]%
        {Jacobs_Zoran_2015}
\bibfield{author}{\bibinfo{person}{Jennifer Jacobs} {and} \bibinfo{person}{Amit
  Zoran}.} \bibinfo{year}{2015}\natexlab{}.
\newblock \showarticletitle{Hybrid Practice in the Kalahari: Design
  Collaboration through Digital Tools and Hunter-Gatherer Craft}. In
  \bibinfo{booktitle}{\emph{Proceedings of the 33rd Annual ACM Conference on
  Human Factors in Computing Systems}} \emph{(\bibinfo{series}{CHI ’15})}.
  \bibinfo{publisher}{Association for Computing Machinery},
  \bibinfo{pages}{619–628}.
\newblock
\showISBNx{978-1-4503-3145-6}
\urldef\tempurl%
\url{https://doi.org/10.1145/2702123.2702362}
\showDOI{\tempurl}


\bibitem[\protect\citeauthoryear{Kazi, Chevalier, Grossman, and
  Fitzmaurice}{Kazi et~al\mbox{.}}{2014}]%
        {Kazi-kitty}
\bibfield{author}{\bibinfo{person}{Rubaiat~Habib Kazi}, \bibinfo{person}{Fanny
  Chevalier}, \bibinfo{person}{Tovi Grossman}, {and} \bibinfo{person}{George
  Fitzmaurice}.} \bibinfo{year}{2014}\natexlab{}.
\newblock \showarticletitle{Kitty: Sketching Dynamic and Interactive
  Illustrations}. In \bibinfo{booktitle}{\emph{Proceedings of the 27th Annual
  ACM Symposium on User Interface Software and Technology}} (Honolulu, Hawaii,
  USA) \emph{(\bibinfo{series}{UIST '14})}. \bibinfo{publisher}{Association for
  Computing Machinery}, \bibinfo{address}{New York, NY, USA},
  \bibinfo{pages}{395–405}.
\newblock
\showISBNx{9781450330695}
\urldef\tempurl%
\url{https://doi.org/10.1145/2642918.2647375}
\showDOI{\tempurl}


\bibitem[\protect\citeauthoryear{Kazi, Chua, Zhao, Davis, and Low}{Kazi
  et~al\mbox{.}}{2011}]%
        {Kazi-sandcanvas}
\bibfield{author}{\bibinfo{person}{Rubaiat~Habib Kazi},
  \bibinfo{person}{Kien~Chuan Chua}, \bibinfo{person}{Shengdong Zhao},
  \bibinfo{person}{Richard Davis}, {and} \bibinfo{person}{Kok-Lim Low}.}
  \bibinfo{year}{2011}\natexlab{}.
\newblock \showarticletitle{SandCanvas: New Possibilities in Sand Animation}.
  In \bibinfo{booktitle}{\emph{CHI '11 Extended Abstracts on Human Factors in
  Computing Systems}} (Vancouver, BC, Canada) \emph{(\bibinfo{series}{CHI EA
  '11})}. \bibinfo{publisher}{Association for Computing Machinery},
  \bibinfo{address}{New York, NY, USA}, \bibinfo{pages}{483}.
\newblock
\showISBNx{9781450302685}
\urldef\tempurl%
\url{https://doi.org/10.1145/1979742.1979562}
\showDOI{\tempurl}


\bibitem[\protect\citeauthoryear{Kazi, Grossman, Cheong, Hashemi, and
  Fitzmaurice}{Kazi et~al\mbox{.}}{2017}]%
        {Kazi-dreamsketch}
\bibfield{author}{\bibinfo{person}{Rubaiat~Habib Kazi}, \bibinfo{person}{Tovi
  Grossman}, \bibinfo{person}{Hyunmin Cheong}, \bibinfo{person}{Ali Hashemi},
  {and} \bibinfo{person}{George Fitzmaurice}.} \bibinfo{year}{2017}\natexlab{}.
\newblock \showarticletitle{DreamSketch: Early Stage 3D Design Explorations
  with Sketching and Generative Design}. In
  \bibinfo{booktitle}{\emph{Proceedings of the 30th Annual ACM Symposium on
  User Interface Software and Technology}} (Qu\'{e}bec City, QC, Canada)
  \emph{(\bibinfo{series}{UIST '17})}. \bibinfo{publisher}{Association for
  Computing Machinery}, \bibinfo{address}{New York, NY, USA},
  \bibinfo{pages}{401–414}.
\newblock
\showISBNx{9781450349819}
\urldef\tempurl%
\url{https://doi.org/10.1145/3126594.3126662}
\showDOI{\tempurl}


\bibitem[\protect\citeauthoryear{Kim, Dontcheva, Li, Bernstein, and
  Steinsapir}{Kim et~al\mbox{.}}{2015}]%
        {Kim-motif}
\bibfield{author}{\bibinfo{person}{Joy Kim}, \bibinfo{person}{Mira Dontcheva},
  \bibinfo{person}{Wilmot Li}, \bibinfo{person}{Michael~S. Bernstein}, {and}
  \bibinfo{person}{Daniela Steinsapir}.} \bibinfo{year}{2015}\natexlab{}.
\newblock \showarticletitle{Motif: Supporting Novice Creativity through Expert
  Patterns}. In \bibinfo{booktitle}{\emph{Proceedings of the 33rd Annual ACM
  Conference on Human Factors in Computing Systems}} (Seoul, Republic of Korea)
  \emph{(\bibinfo{series}{CHI '15})}. \bibinfo{publisher}{Association for
  Computing Machinery}, \bibinfo{address}{New York, NY, USA},
  \bibinfo{pages}{1211–1220}.
\newblock
\showISBNx{9781450331456}
\urldef\tempurl%
\url{https://doi.org/10.1145/2702123.2702507}
\showDOI{\tempurl}


\bibitem[\protect\citeauthoryear{Klemmer, Hartmann, and Takayama}{Klemmer
  et~al\mbox{.}}{2006}]%
        {Klemmer_Hartmann_Takayama_2006}
\bibfield{author}{\bibinfo{person}{Scott~R. Klemmer}, \bibinfo{person}{Björn
  Hartmann}, {and} \bibinfo{person}{Leila Takayama}.}
  \bibinfo{year}{2006}\natexlab{}.
\newblock \showarticletitle{How bodies matter: five themes for interaction
  design}. In \bibinfo{booktitle}{\emph{Proceedings of the 6th conference on
  Designing Interactive systems}} \emph{(\bibinfo{series}{DIS ’06})}.
  \bibinfo{publisher}{Association for Computing Machinery},
  \bibinfo{pages}{140–149}.
\newblock
\showISBNx{978-1-59593-367-6}
\urldef\tempurl%
\url{https://doi.org/10.1145/1142405.1142429}
\showDOI{\tempurl}


\bibitem[\protect\citeauthoryear{Klokmose, Eagan, Baader, Mackay, and
  Beaudouin-Lafon}{Klokmose et~al\mbox{.}}{2015}]%
        {Klokmose_Eagan_Baader_Mackay_Beaudouin-Lafon_2015}
\bibfield{author}{\bibinfo{person}{Clemens~N. Klokmose},
  \bibinfo{person}{James~R. Eagan}, \bibinfo{person}{Siemen Baader},
  \bibinfo{person}{Wendy Mackay}, {and} \bibinfo{person}{Michel
  Beaudouin-Lafon}.} \bibinfo{year}{2015}\natexlab{}.
\newblock \showarticletitle{Webstrates: Shareable Dynamic Media}. In
  \bibinfo{booktitle}{\emph{Proceedings of the 28th Annual ACM Symposium on
  User Interface Software \& Technology}} \emph{(\bibinfo{series}{UIST
  ’15})}. \bibinfo{publisher}{Association for Computing Machinery},
  \bibinfo{pages}{280–290}.
\newblock
\showISBNx{978-1-4503-3779-3}
\urldef\tempurl%
\url{https://doi.org/10.1145/2807442.2807446}
\showDOI{\tempurl}


\bibitem[\protect\citeauthoryear{Klokmose, Remy, Kristensen, Bagge,
  Beaudouin-Lafon, and Mackay}{Klokmose et~al\mbox{.}}{2019}]%
        {Klokmose_Remy_Kristensen_Bagge_Beaudouin-Lafon_Mackay_2019}
\bibfield{author}{\bibinfo{person}{Clemens~N. Klokmose},
  \bibinfo{person}{Christian Remy}, \bibinfo{person}{Janus~Bager Kristensen},
  \bibinfo{person}{Rolf Bagge}, \bibinfo{person}{Michel Beaudouin-Lafon}, {and}
  \bibinfo{person}{Wendy Mackay}.} \bibinfo{year}{2019}\natexlab{}.
\newblock \showarticletitle{Videostrates: Collaborative, Distributed and
  Programmable Video Manipulation}. In \bibinfo{booktitle}{\emph{Proceedings of
  the 32nd Annual ACM Symposium on User Interface Software and Technology}}
  \emph{(\bibinfo{series}{UIST ’19})}. \bibinfo{publisher}{Association for
  Computing Machinery}, \bibinfo{pages}{233–247}.
\newblock
\showISBNx{978-1-4503-6816-2}
\urldef\tempurl%
\url{https://doi.org/10.1145/3332165.3347912}
\showDOI{\tempurl}


\bibitem[\protect\citeauthoryear{Ko, Abraham, Beckwith, Blackwell, Burnett,
  Erwig, Scaffidi, Lawrance, Lieberman, Myers, and et~al.}{Ko
  et~al\mbox{.}}{2011}]%
        {Ko_Abraham}
\bibfield{author}{\bibinfo{person}{Amy~J. Ko}, \bibinfo{person}{Robin Abraham},
  \bibinfo{person}{Laura Beckwith}, \bibinfo{person}{Alan Blackwell},
  \bibinfo{person}{Margaret Burnett}, \bibinfo{person}{Martin Erwig},
  \bibinfo{person}{Chris Scaffidi}, \bibinfo{person}{Joseph Lawrance},
  \bibinfo{person}{Henry Lieberman}, \bibinfo{person}{Brad Myers}, {and}
  \bibinfo{person}{et al.}} \bibinfo{year}{2011}\natexlab{}.
\newblock \showarticletitle{The state of the art in end-user software
  engineering}.
\newblock \bibinfo{journal}{\emph{Comput. Surveys}} \bibinfo{volume}{43},
  \bibinfo{number}{3} (\bibinfo{date}{Apr} \bibinfo{year}{2011}),
  \bibinfo{pages}{21:1–21:44}.
\newblock
\showISSN{0360-0300}
\urldef\tempurl%
\url{https://doi.org/10.1145/1922649.1922658}
\showDOI{\tempurl}


\bibitem[\protect\citeauthoryear{Leiva, Maudet, Mackay, and
  Beaudouin-Lafon}{Leiva et~al\mbox{.}}{2019}]%
        {enact}
\bibfield{author}{\bibinfo{person}{Germ\'{a}n Leiva}, \bibinfo{person}{Nolwenn
  Maudet}, \bibinfo{person}{Wendy Mackay}, {and} \bibinfo{person}{Michel
  Beaudouin-Lafon}.} \bibinfo{year}{2019}\natexlab{}.
\newblock \showarticletitle{Enact: Reducing Designer–Developer Breakdowns
  When Prototyping Custom Interactions}.
\newblock \bibinfo{journal}{\emph{ACM Trans. Comput.-Hum. Interact.}}
  \bibinfo{volume}{26}, \bibinfo{number}{3}, Article \bibinfo{articleno}{19}
  (\bibinfo{date}{May} \bibinfo{year}{2019}), \bibinfo{numpages}{48}~pages.
\newblock
\showISSN{1073-0516}
\urldef\tempurl%
\url{https://doi.org/10.1145/3310276}
\showDOI{\tempurl}


\bibitem[\protect\citeauthoryear{Leung and Lara}{Leung and Lara}{2015}]%
        {Leung-greasepencil}
\bibfield{author}{\bibinfo{person}{Joshua Leung} {and}
  \bibinfo{person}{Daniel~M. Lara}.} \bibinfo{year}{2015}\natexlab{}.
\newblock \showarticletitle{Grease Pencil: Integrating Animated Freehand
  Drawings into 3D Production Environments}. In
  \bibinfo{booktitle}{\emph{SIGGRAPH Asia 2015 Technical Briefs}} (Kobe, Japan)
  \emph{(\bibinfo{series}{SA '15})}. \bibinfo{publisher}{Association for
  Computing Machinery}, \bibinfo{address}{New York, NY, USA}, Article
  \bibinfo{articleno}{16}, \bibinfo{numpages}{4}~pages.
\newblock
\showISBNx{9781450339308}
\urldef\tempurl%
\url{https://doi.org/10.1145/2820903.2820924}
\showDOI{\tempurl}


\bibitem[\protect\citeauthoryear{Levin}{Levin}{2003}]%
        {levin2003essay}
\bibfield{author}{\bibinfo{person}{Golan Levin}.}
  \bibinfo{year}{2003}\natexlab{}.
\newblock \bibinfo{booktitle}{\emph{Essay for creative code}}.
\newblock
\urldef\tempurl%
\url{http://www.flong.com/texts/essays/essay\_creative\_code}
\showURL{%
\tempurl}


\bibitem[\protect\citeauthoryear{Levin}{Levin}{2015}]%
        {Golan}
\bibfield{author}{\bibinfo{person}{Golan Levin}.}
  \bibinfo{year}{2015}\natexlab{}.
\newblock \bibinfo{booktitle}{\emph{For Us, By Us}}.
\newblock
\urldef\tempurl%
\url{http://www.flong.com/texts/essays/for-us-by-us/}
\showURL{%
\tempurl}


\bibitem[\protect\citeauthoryear{Li, Brandt, Mech, Agrawala, and Jacobs}{Li
  et~al\mbox{.}}{2020}]%
        {Li-ddb}
\bibfield{author}{\bibinfo{person}{Jingyi Li}, \bibinfo{person}{Joel Brandt},
  \bibinfo{person}{Radom\'{\i}r Mech}, \bibinfo{person}{Maneesh Agrawala},
  {and} \bibinfo{person}{Jennifer Jacobs}.} \bibinfo{year}{2020}\natexlab{}.
\newblock \showarticletitle{Supporting Visual Artists in Programming through
  Direct Inspection and Control of Program Execution}. In
  \bibinfo{booktitle}{\emph{Proceedings of the 2020 CHI Conference on Human
  Factors in Computing Systems}} (Honolulu, HI, USA)
  \emph{(\bibinfo{series}{CHI '20})}. \bibinfo{publisher}{Association for
  Computing Machinery}, \bibinfo{address}{New York, NY, USA},
  \bibinfo{pages}{1–12}.
\newblock
\showISBNx{9781450367080}
\urldef\tempurl%
\url{https://doi.org/10.1145/3313831.3376765}
\showDOI{\tempurl}


\bibitem[\protect\citeauthoryear{Lieberman, Patern{\`o}, Klann, and
  Wulf}{Lieberman et~al\mbox{.}}{2006}]%
        {lieberman2006end}
\bibfield{author}{\bibinfo{person}{Henry Lieberman}, \bibinfo{person}{Fabio
  Patern{\`o}}, \bibinfo{person}{Markus Klann}, {and} \bibinfo{person}{Volker
  Wulf}.} \bibinfo{year}{2006}\natexlab{}.
\newblock \showarticletitle{End-user development: An emerging paradigm}.
\newblock In \bibinfo{booktitle}{\emph{End user development}}.
  \bibinfo{publisher}{Springer}, \bibinfo{pages}{1--8}.
\newblock


\bibitem[\protect\citeauthoryear{Lieberman}{Lieberman}{2014}]%
        {ofbook}
\bibfield{author}{\bibinfo{person}{Zach Lieberman}.}
  \bibinfo{year}{2014}\natexlab{}.
\newblock \bibinfo{booktitle}{\emph{ofBook, a collaboratively written book
  about openFrameworks}}.
\newblock
\newblock
\shownote{\url{http://openframeworks.cc/ofBook/chapters/foreword.html}.}


\bibitem[\protect\citeauthoryear{Malloch, Garcia, Wanderley, Mackay,
  Beaudouin-Lafon, and Huot}{Malloch et~al\mbox{.}}{2019}]%
        {WorkbenchMusic2019}
\bibfield{author}{\bibinfo{person}{Joseph Malloch},
  \bibinfo{person}{J{\'e}r{\'e}mie Garcia}, \bibinfo{person}{Marcelo~M.
  Wanderley}, \bibinfo{person}{Wendy~E. Mackay}, \bibinfo{person}{Michel
  Beaudouin-Lafon}, {and} \bibinfo{person}{St{\'e}phane Huot}.}
  \bibinfo{year}{2019}\natexlab{}.
\newblock \bibinfo{booktitle}{\emph{A Design Workbench for Interactive Music
  Systems}}.
\newblock \bibinfo{publisher}{Springer International Publishing},
  \bibinfo{address}{Cham}, \bibinfo{pages}{23--40}.
\newblock
\showISBNx{978-3-319-92069-6}
\urldef\tempurl%
\url{https://doi.org/10.1007/978-3-319-92069-6_2}
\showDOI{\tempurl}


\bibitem[\protect\citeauthoryear{Margolis and Fisher}{Margolis and
  Fisher}{2002}]%
        {Margolis_Fisher_2002}
\bibfield{author}{\bibinfo{person}{J. Margolis} {and} \bibinfo{person}{A.
  Fisher}.} \bibinfo{year}{2002}\natexlab{}.
\newblock \bibinfo{booktitle}{\emph{Unlocking the Clubhouse: Women in
  Computing}}.
\newblock \bibinfo{publisher}{MIT Press}.
\newblock
\showISBNx{978-0-262-63269-0}
\urldef\tempurl%
\url{https://books.google.com/books?id=StwGQw45YoEC}
\showURL{%
\tempurl}


\bibitem[\protect\citeauthoryear{Matejka, Glueck, Bradner, Hashemi, Grossman,
  and Fitzmaurice}{Matejka et~al\mbox{.}}{2018}]%
        {Matejka_Glueck_Bradner_Hashemi_Grossman_Fitzmaurice_2018}
\bibfield{author}{\bibinfo{person}{Justin Matejka}, \bibinfo{person}{Michael
  Glueck}, \bibinfo{person}{Erin Bradner}, \bibinfo{person}{Ali Hashemi},
  \bibinfo{person}{Tovi Grossman}, {and} \bibinfo{person}{George Fitzmaurice}.}
  \bibinfo{year}{2018}\natexlab{}.
\newblock \showarticletitle{Dream Lens: Exploration and Visualization of
  Large-Scale Generative Design Datasets}. In
  \bibinfo{booktitle}{\emph{Proceedings of the 2018 CHI Conference on Human
  Factors in Computing Systems - CHI ’18}}. \bibinfo{publisher}{ACM Press},
  \bibinfo{pages}{1–12}.
\newblock
\showISBNx{978-1-4503-5620-6}
\urldef\tempurl%
\url{https://doi.org/10.1145/3173574.3173943}
\showDOI{\tempurl}


\bibitem[\protect\citeauthoryear{McCarthy and Turner}{McCarthy and
  Turner}{[n.d.]a}]%
        {p5.js}
\bibfield{author}{\bibinfo{person}{L. McCarthy} {and} \bibinfo{person}{M.
  Turner}.} \bibinfo{year}{[n.d.]}\natexlab{a}.
\newblock \bibinfo{booktitle}{\emph{p5.js}}.
\newblock
\newblock
\shownote{\url{https://p5js.org/}.}


\bibitem[\protect\citeauthoryear{McCarthy and Turner}{McCarthy and
  Turner}{[n.d.]b}]%
        {p5.js-community}
\bibfield{author}{\bibinfo{person}{L. McCarthy} {and} \bibinfo{person}{M.
  Turner}.} \bibinfo{year}{[n.d.]}\natexlab{b}.
\newblock \bibinfo{booktitle}{\emph{p5.js Community Statement}}.
\newblock
\newblock
\shownote{\url{https://p5js.org/community/}.}


\bibitem[\protect\citeauthoryear{McCullough}{McCullough}{1996}]%
        {McCullough}
\bibfield{author}{\bibinfo{person}{Malcom McCullough}.}
  \bibinfo{year}{1996}\natexlab{}.
\newblock \bibinfo{booktitle}{\emph{Abstracting craft: the practiced digital
  hand}}.
\newblock \bibinfo{publisher}{MIT Press}.
\newblock


\bibitem[\protect\citeauthoryear{McDirmid}{McDirmid}{2016}]%
        {McDirmid-apx}
\bibfield{author}{\bibinfo{person}{Sean McDirmid}.}
  \bibinfo{year}{2016}\natexlab{}.
\newblock \bibinfo{title}{A Live Programming Experience}.
\newblock
  \bibinfo{howpublished}{\url{https://www.youtube.com/watch?v=bnqkglrSqrg}}.
\newblock


\bibitem[\protect\citeauthoryear{Mordvintsev, Olah, and Tyka}{Mordvintsev
  et~al\mbox{.}}{2015}]%
        {mordvintsev2015deepdream}
\bibfield{author}{\bibinfo{person}{Alexander Mordvintsev},
  \bibinfo{person}{Christopher Olah}, {and} \bibinfo{person}{Mike Tyka}.}
  \bibinfo{year}{2015}\natexlab{}.
\newblock \showarticletitle{Deepdream-a code example for visualizing neural
  networks}.
\newblock \bibinfo{journal}{\emph{Google Research}} \bibinfo{volume}{2},
  \bibinfo{number}{5} (\bibinfo{year}{2015}).
\newblock


\bibitem[\protect\citeauthoryear{Muller and Kuhn}{Muller and Kuhn}{1993}]%
        {muller1993participatory}
\bibfield{author}{\bibinfo{person}{Michael~J Muller} {and}
  \bibinfo{person}{Sarah Kuhn}.} \bibinfo{year}{1993}\natexlab{}.
\newblock \showarticletitle{Participatory design}.
\newblock \bibinfo{journal}{\emph{Commun. ACM}} \bibinfo{volume}{36},
  \bibinfo{number}{6} (\bibinfo{year}{1993}), \bibinfo{pages}{24--28}.
\newblock


\bibitem[\protect\citeauthoryear{Mumford}{Mumford}{1952}]%
        {Mumford}
\bibfield{author}{\bibinfo{person}{Lewsi Mumford}.}
  \bibinfo{year}{1952}\natexlab{}.
\newblock \bibinfo{booktitle}{\emph{Art and Technics}}.
\newblock \bibinfo{publisher}{Columbia University Press}.
\newblock


\bibitem[\protect\citeauthoryear{Myers, Park, Nakano, Mueller, and Ko}{Myers
  et~al\mbox{.}}{2008}]%
        {myers2008designers}
\bibfield{author}{\bibinfo{person}{Brad Myers}, \bibinfo{person}{Sun~Young
  Park}, \bibinfo{person}{Yoko Nakano}, \bibinfo{person}{Greg Mueller}, {and}
  \bibinfo{person}{Andrew Ko}.} \bibinfo{year}{2008}\natexlab{}.
\newblock \showarticletitle{How designers design and program interactive
  behaviors}. In \bibinfo{booktitle}{\emph{2008 IEEE Symposium on Visual
  Languages and Human-Centric Computing}}. IEEE, \bibinfo{pages}{177--184}.
\newblock


\bibitem[\protect\citeauthoryear{Myers, Lai, Le, Yoon, Faulring, and
  Brandt}{Myers et~al\mbox{.}}{2015}]%
        {Myers-undo}
\bibfield{author}{\bibinfo{person}{Brad~A. Myers}, \bibinfo{person}{Ashley
  Lai}, \bibinfo{person}{Tam~Minh Le}, \bibinfo{person}{YoungSeok Yoon},
  \bibinfo{person}{Andrew Faulring}, {and} \bibinfo{person}{Joel Brandt}.}
  \bibinfo{year}{2015}\natexlab{}.
\newblock \showarticletitle{Selective Undo Support for Painting Applications}.
  In \bibinfo{booktitle}{\emph{Proceedings of the 33rd Annual ACM Conference on
  Human Factors in Computing Systems}} (Seoul, Republic of Korea)
  \emph{(\bibinfo{series}{CHI '15})}. \bibinfo{publisher}{Association for
  Computing Machinery}, \bibinfo{address}{New York, NY, USA},
  \bibinfo{pages}{4227–4236}.
\newblock
\showISBNx{9781450331456}
\urldef\tempurl%
\url{https://doi.org/10.1145/2702123.2702543}
\showDOI{\tempurl}


\bibitem[\protect\citeauthoryear{Needleman}{Needleman}{1979}]%
        {Needleman}
\bibfield{author}{\bibinfo{person}{Carla Needleman}.}
  \bibinfo{year}{1979}\natexlab{}.
\newblock \bibinfo{booktitle}{\emph{The work of craft: an inquiry into the
  nature of crafts and craftsmanship}}.
\newblock \bibinfo{publisher}{Arkana}.
\newblock


\bibitem[\protect\citeauthoryear{Oney, Myers, and Brandt}{Oney
  et~al\mbox{.}}{2014}]%
        {Oney_Myers_Brandt_2014}
\bibfield{author}{\bibinfo{person}{Stephen Oney}, \bibinfo{person}{Brad Myers},
  {and} \bibinfo{person}{Joel Brandt}.} \bibinfo{year}{2014}\natexlab{}.
\newblock \showarticletitle{InterState: a language and environment for
  expressing interface behavior}. In \bibinfo{booktitle}{\emph{Proceedings of
  the 27th annual ACM symposium on User interface software and technology}}
  \emph{(\bibinfo{series}{UIST ’14})}. \bibinfo{publisher}{Association for
  Computing Machinery}, \bibinfo{pages}{263–272}.
\newblock
\showISBNx{978-1-4503-3069-5}
\urldef\tempurl%
\url{https://doi.org/10.1145/2642918.2647358}
\showDOI{\tempurl}


\bibitem[\protect\citeauthoryear{Papert}{Papert}{1980}]%
        {Papert-mindstorms}
\bibfield{author}{\bibinfo{person}{Seymour Papert}.}
  \bibinfo{year}{1980}\natexlab{}.
\newblock \bibinfo{booktitle}{\emph{Mindstorms: Children, Computers, and
  Powerful Ideas}}.
\newblock \bibinfo{publisher}{Basic Books, Inc.}, \bibinfo{address}{USA}.
\newblock
\showISBNx{0465046274}


\bibitem[\protect\citeauthoryear{Peek and Gershenfeld}{Peek and
  Gershenfeld}{2018}]%
        {peek2018mods}
\bibfield{author}{\bibinfo{person}{Nadya Peek} {and} \bibinfo{person}{Neil
  Gershenfeld}.} \bibinfo{year}{2018}\natexlab{}.
\newblock \showarticletitle{Mods: Browser-based rapid prototyping workflow
  composition}.
\newblock  (\bibinfo{year}{2018}).
\newblock


\bibitem[\protect\citeauthoryear{Reas and Fry}{Reas and Fry}{2004}]%
        {processing}
\bibfield{author}{\bibinfo{person}{C. Reas} {and} \bibinfo{person}{B. Fry}.}
  \bibinfo{year}{2004}\natexlab{}.
\newblock \bibinfo{booktitle}{\emph{Processing}}.
\newblock
\newblock
\shownote{\url{http://processing.org}.}


\bibitem[\protect\citeauthoryear{Reas and Fry}{Reas and Fry}{2016}]%
        {processing_overview}
\bibfield{author}{\bibinfo{person}{C. Reas} {and} \bibinfo{person}{B. Fry}.}
  \bibinfo{year}{2016}\natexlab{}.
\newblock \bibinfo{booktitle}{\emph{Processing Overview}}.
\newblock
\newblock
\shownote{\url{http://processing.org/overview}.}


\bibitem[\protect\citeauthoryear{Resnick and Silverman}{Resnick and
  Silverman}{2005}]%
        {Resnick_Silverman_2005}
\bibfield{author}{\bibinfo{person}{Mitchel Resnick} {and}
  \bibinfo{person}{Brian Silverman}.} \bibinfo{year}{2005}\natexlab{}.
\newblock \showarticletitle{Some reflections on designing construction kits for
  kids}. In \bibinfo{booktitle}{\emph{Proceedings of the 2005 conference on
  Interaction design and children}} \emph{(\bibinfo{series}{IDC ’05})}.
  \bibinfo{publisher}{Association for Computing Machinery},
  \bibinfo{pages}{117–122}.
\newblock
\showISBNx{978-1-59593-096-5}
\urldef\tempurl%
\url{https://doi.org/10.1145/1109540.1109556}
\showDOI{\tempurl}


\bibitem[\protect\citeauthoryear{Roque}{Roque}{2016}]%
        {roque2016family}
\bibfield{author}{\bibinfo{person}{Ricarose~Vallarta Roque}.}
  \bibinfo{year}{2016}\natexlab{}.
\newblock \emph{\bibinfo{title}{Family creative learning: designing structures
  to engage kids and parents as computational creators}}.
\newblock \bibinfo{thesistype}{Ph.D. Dissertation}.
  \bibinfo{school}{Massachusetts Institute of Technology}.
\newblock


\bibitem[\protect\citeauthoryear{Runway~AI}{Runway~AI}{2020}]%
        {RunwayML}
\bibfield{author}{\bibinfo{person}{Inc. Runway~AI}.}
  \bibinfo{year}{2020}\natexlab{}.
\newblock \bibinfo{booktitle}{\emph{RunwayML}}.
\newblock
\newblock
\shownote{\url{https://runwayml.com/}.}


\bibitem[\protect\citeauthoryear{Savage, Follmer, Li, and Hartmann}{Savage
  et~al\mbox{.}}{2015}]%
        {savage-mm-10.1145/2807442.2807508}
\bibfield{author}{\bibinfo{person}{Valkyrie Savage}, \bibinfo{person}{Sean
  Follmer}, \bibinfo{person}{Jingyi Li}, {and} \bibinfo{person}{Bj\"{o}rn
  Hartmann}.} \bibinfo{year}{2015}\natexlab{}.
\newblock \showarticletitle{Makers' Marks: Physical Markup for Designing and
  Fabricating Functional Objects}. In \bibinfo{booktitle}{\emph{Proceedings of
  the 28th Annual ACM Symposium on User Interface Software \& Technology}}
  (Charlotte, NC, USA) \emph{(\bibinfo{series}{UIST '15})}.
  \bibinfo{publisher}{Association for Computing Machinery},
  \bibinfo{address}{New York, NY, USA}, \bibinfo{pages}{103–108}.
\newblock
\showISBNx{9781450337793}
\urldef\tempurl%
\url{https://doi.org/10.1145/2807442.2807508}
\showDOI{\tempurl}


\bibitem[\protect\citeauthoryear{Schachman}{Schachman}{2012}]%
        {schacman10.1145/2384592.2384594}
\bibfield{author}{\bibinfo{person}{Toby Schachman}.}
  \bibinfo{year}{2012}\natexlab{}.
\newblock \showarticletitle{Alternative Programming Interfaces for Alternative
  Programmers}. In \bibinfo{booktitle}{\emph{Proceedings of the ACM
  International Symposium on New Ideas, New Paradigms, and Reflections on
  Programming and Software}} (Tucson, Arizona, USA)
  \emph{(\bibinfo{series}{Onward! 2012})}. \bibinfo{publisher}{Association for
  Computing Machinery}, \bibinfo{address}{New York, NY, USA},
  \bibinfo{pages}{1–10}.
\newblock
\showISBNx{9781450315623}
\urldef\tempurl%
\url{https://doi.org/10.1145/2384592.2384594}
\showDOI{\tempurl}


\bibitem[\protect\citeauthoryear{Schon}{Schon}{[n.d.]}]%
        {Schon}
\bibfield{author}{\bibinfo{person}{Donald~A Schon}.}
  \bibinfo{year}{[n.d.]}\natexlab{}.
\newblock \showarticletitle{Designing as reflective conversation with the
  materials of a design situation}.
\newblock  (\bibinfo{year}{[n.\,d.]}), \bibinfo{pages}{17}.
\newblock


\bibitem[\protect\citeauthoryear{Semmo, Isenberg, and D\"{o}llner}{Semmo
  et~al\mbox{.}}{2017}]%
        {Semmo}
\bibfield{author}{\bibinfo{person}{Amir Semmo}, \bibinfo{person}{Tobias
  Isenberg}, {and} \bibinfo{person}{J\"{u}rgen D\"{o}llner}.}
  \bibinfo{year}{2017}\natexlab{}.
\newblock \showarticletitle{Neural Style Transfer: A Paradigm Shift for
  Image-Based Artistic Rendering?}. In \bibinfo{booktitle}{\emph{Proceedings of
  the Symposium on Non-Photorealistic Animation and Rendering}} (Los Angeles,
  California) \emph{(\bibinfo{series}{NPAR '17})}.
  \bibinfo{publisher}{Association for Computing Machinery},
  \bibinfo{address}{New York, NY, USA}, Article \bibinfo{articleno}{5},
  \bibinfo{numpages}{13}~pages.
\newblock
\showISBNx{9781450350815}
\urldef\tempurl%
\url{https://doi.org/10.1145/3092919.3092920}
\showDOI{\tempurl}


\bibitem[\protect\citeauthoryear{Shimizu, Fisher, Paris, and
  Fatahalian}{Shimizu et~al\mbox{.}}{2019}]%
        {layers-10.20380/GI2019.16}
\bibfield{author}{\bibinfo{person}{Evan Shimizu}, \bibinfo{person}{Matt
  Fisher}, \bibinfo{person}{Sylvain Paris}, {and} \bibinfo{person}{Kayvon
  Fatahalian}.} \bibinfo{year}{2019}\natexlab{}.
\newblock \showarticletitle{Finding Layers Using Hover Visualizations}. In
  \bibinfo{booktitle}{\emph{Proceedings of the 45th Graphics Interface
  Conference on Proceedings of Graphics Interface 2019}} (Kingston, Canada)
  \emph{(\bibinfo{series}{GI'19})}. \bibinfo{publisher}{Canadian Human-Computer
  Communications Society}, \bibinfo{address}{Waterloo, CAN}, Article
  \bibinfo{articleno}{16}, \bibinfo{numpages}{9}~pages.
\newblock
\showISBNx{9780994786845}
\urldef\tempurl%
\url{https://doi.org/10.20380/GI2019.16}
\showDOI{\tempurl}


\bibitem[\protect\citeauthoryear{Shneiderman}{Shneiderman}{2002}]%
        {Shneiderman-csts}
\bibfield{author}{\bibinfo{person}{Ben Shneiderman}.}
  \bibinfo{year}{2002}\natexlab{}.
\newblock \showarticletitle{Creativity Support Tools}.
\newblock \bibinfo{journal}{\emph{Commun. ACM}} \bibinfo{volume}{45},
  \bibinfo{number}{10} (\bibinfo{date}{Oct.} \bibinfo{year}{2002}),
  \bibinfo{pages}{116–120}.
\newblock
\showISSN{0001-0782}
\urldef\tempurl%
\url{https://doi.org/10.1145/570907.570945}
\showDOI{\tempurl}


\bibitem[\protect\citeauthoryear{Shneiderman}{Shneiderman}{2007}]%
        {Shneiderman_2007}
\bibfield{author}{\bibinfo{person}{Ben Shneiderman}.}
  \bibinfo{year}{2007}\natexlab{}.
\newblock \showarticletitle{Creativity support tools: accelerating discovery
  and innovation}.
\newblock \bibinfo{journal}{\emph{Commun. ACM}} \bibinfo{volume}{50},
  \bibinfo{number}{12} (\bibinfo{date}{Dec} \bibinfo{year}{2007}),
  \bibinfo{pages}{20–32}.
\newblock
\showISSN{00010782}
\urldef\tempurl%
\url{https://doi.org/10.1145/1323688.1323689}
\showDOI{\tempurl}


\bibitem[\protect\citeauthoryear{Simo-Serra, Iizuka, and Ishikawa}{Simo-Serra
  et~al\mbox{.}}{2018}]%
        {Simo-Serra-2}
\bibfield{author}{\bibinfo{person}{Edgar Simo-Serra}, \bibinfo{person}{Satoshi
  Iizuka}, {and} \bibinfo{person}{Hiroshi Ishikawa}.}
  \bibinfo{year}{2018}\natexlab{}.
\newblock \showarticletitle{{Mastering Sketching: Adversarial Augmentation for
  Structured Prediction}}.
\newblock \bibinfo{journal}{\emph{Transactions on Graphics (Presented at
  SIGGRAPH)}} \bibinfo{volume}{37}, \bibinfo{number}{1} (\bibinfo{year}{2018}).
\newblock


\bibitem[\protect\citeauthoryear{Simon}{Simon}{2019}]%
        {artbreeder}
\bibfield{author}{\bibinfo{person}{Joel Simon}.}
  \bibinfo{year}{2019}\natexlab{}.
\newblock \bibinfo{booktitle}{\emph{ArtBreeder}}.
\newblock
\urldef\tempurl%
\url{https://www.artbreeder.com/}
\showURL{%
\tempurl}


\bibitem[\protect\citeauthoryear{Thiel, Singh, and Balakrishnan}{Thiel
  et~al\mbox{.}}{2011}]%
        {Thiel-elasticurves}
\bibfield{author}{\bibinfo{person}{Yannick Thiel}, \bibinfo{person}{Karan
  Singh}, {and} \bibinfo{person}{Ravin Balakrishnan}.}
  \bibinfo{year}{2011}\natexlab{}.
\newblock \showarticletitle{Elasticurves: Exploiting Stroke Dynamics and
  Inertia for the Real-Time Neatening of Sketched 2D Curves}. In
  \bibinfo{booktitle}{\emph{Proceedings of the 24th Annual ACM Symposium on
  User Interface Software and Technology}} (Santa Barbara, California, USA)
  \emph{(\bibinfo{series}{UIST '11})}. \bibinfo{publisher}{Association for
  Computing Machinery}, \bibinfo{address}{New York, NY, USA},
  \bibinfo{pages}{383–392}.
\newblock
\showISBNx{9781450307161}
\urldef\tempurl%
\url{https://doi.org/10.1145/2047196.2047246}
\showDOI{\tempurl}


\bibitem[\protect\citeauthoryear{Torres, J\"{o}rke, Hill, and Paulos}{Torres
  et~al\mbox{.}}{2019}]%
        {torres10.1145/3325480.3325498}
\bibfield{author}{\bibinfo{person}{Cesar Torres}, \bibinfo{person}{Matthew
  J\"{o}rke}, \bibinfo{person}{Emily Hill}, {and} \bibinfo{person}{Eric
  Paulos}.} \bibinfo{year}{2019}\natexlab{}.
\newblock \showarticletitle{Hybrid Microgenetic Analysis: Using Activity
  Codebooks to Identify and Characterize Creative Process}. In
  \bibinfo{booktitle}{\emph{Proceedings of the 2019 on Creativity and
  Cognition}} (San Diego, CA, USA) \emph{(\bibinfo{series}{C\&C '19})}.
  \bibinfo{publisher}{Association for Computing Machinery},
  \bibinfo{address}{New York, NY, USA}, \bibinfo{pages}{2–14}.
\newblock
\showISBNx{9781450359177}
\urldef\tempurl%
\url{https://doi.org/10.1145/3325480.3325498}
\showDOI{\tempurl}


\bibitem[\protect\citeauthoryear{Tran~O'Leary and Peek}{Tran~O'Leary and
  Peek}{2019}]%
        {tran-mom-10.1145/3332167.3356897}
\bibfield{author}{\bibinfo{person}{Jasper Tran~O'Leary} {and}
  \bibinfo{person}{Nadya Peek}.} \bibinfo{year}{2019}\natexlab{}.
\newblock \showarticletitle{Machine-o-Matic: A Programming Environment for
  Prototyping Digital Fabrication Workflows}. In \bibinfo{booktitle}{\emph{The
  Adjunct Publication of the 32nd Annual ACM Symposium on User Interface
  Software and Technology}} (New Orleans, LA, USA) \emph{(\bibinfo{series}{UIST
  '19})}. \bibinfo{publisher}{Association for Computing Machinery},
  \bibinfo{address}{New York, NY, USA}, \bibinfo{pages}{134–136}.
\newblock
\showISBNx{9781450368179}
\urldef\tempurl%
\url{https://doi.org/10.1145/3332167.3356897}
\showDOI{\tempurl}


\bibitem[\protect\citeauthoryear{Tsandilas, Grammatikou, and Huot}{Tsandilas
  et~al\mbox{.}}{2015}]%
        {bricosketch-10.1145/2817721.2817729}
\bibfield{author}{\bibinfo{person}{Theophanis Tsandilas},
  \bibinfo{person}{Magdalini Grammatikou}, {and} \bibinfo{person}{St\'{e}phane
  Huot}.} \bibinfo{year}{2015}\natexlab{}.
\newblock \showarticletitle{BricoSketch: Mixing Paper and Computer Drawing
  Tools in Professional Illustration}. In \bibinfo{booktitle}{\emph{Proceedings
  of the 2015 International Conference on Interactive Tabletops \& Surfaces}}
  (Madeira, Portugal) \emph{(\bibinfo{series}{ITS '15})}.
  \bibinfo{publisher}{Association for Computing Machinery},
  \bibinfo{address}{New York, NY, USA}, \bibinfo{pages}{127–136}.
\newblock
\showISBNx{9781450338998}
\urldef\tempurl%
\url{https://doi.org/10.1145/2817721.2817729}
\showDOI{\tempurl}


\bibitem[\protect\citeauthoryear{Tversky}{Tversky}{2019}]%
        {Tversky}
\bibfield{author}{\bibinfo{person}{Barbara Tversky}.}
  \bibinfo{year}{2019}\natexlab{}.
\newblock \bibinfo{booktitle}{\emph{Mind in motion}}.
\newblock \bibinfo{publisher}{Basic Books}.
\newblock


\bibitem[\protect\citeauthoryear{Victor}{Victor}{2011}]%
        {victor2011dynamic}
\bibfield{author}{\bibinfo{person}{Bret Victor}.}
  \bibinfo{year}{2011}\natexlab{}.
\newblock \bibinfo{booktitle}{\emph{Dynamic Pictures}}.
\newblock
\urldef\tempurl%
\url{http://worrydream.com/DynamicPicturesMotivation}
\showURL{%
\tempurl}


\bibitem[\protect\citeauthoryear{Victor}{Victor}{2018}]%
        {dynamicland}
\bibfield{author}{\bibinfo{person}{Bret Victor}.}
  \bibinfo{year}{2018}\natexlab{}.
\newblock \bibinfo{title}{Dynamicland}.
\newblock
\newblock
\urldef\tempurl%
\url{https://dynamicland.org}
\showURL{%
\tempurl}


\bibitem[\protect\citeauthoryear{vvvv group}{vvvv group}{2017}]%
        {vvvv}
\bibfield{author}{\bibinfo{person}{vvvv group}.}
  \bibinfo{year}{2017}\natexlab{}.
\newblock \bibinfo{booktitle}{\emph{vvvv}}.
\newblock
\newblock
\shownote{\url{https://vvvv.org/}.}


\bibitem[\protect\citeauthoryear{Watz}{Watz}{2012}]%
        {watz-eyeo}
\bibfield{author}{\bibinfo{person}{Marius Watz}.}
  \bibinfo{year}{2012}\natexlab{}.
\newblock \bibinfo{booktitle}{\emph{Algorithm Critique and Computational
  Aesthetics}}.
\newblock Vimeo.
\newblock
\urldef\tempurl%
\url{https://vimeo.com/46903693}
\showURL{%
\tempurl}


\bibitem[\protect\citeauthoryear{Wessel and Wright}{Wessel and Wright}{2001}]%
        {Wessel2001}
\bibfield{author}{\bibinfo{person}{David Wessel} {and} \bibinfo{person}{Matthew
  Wright}.} \bibinfo{year}{2001}\natexlab{}.
\newblock \showarticletitle{Problems and Prospects for Intimate Musical Control
  of Computers}. In \bibinfo{booktitle}{\emph{Proceedings of the 2001
  Conference on New Interfaces for Musical Expression}} (Seattle, Washington)
  \emph{(\bibinfo{series}{NIME '01})}. \bibinfo{publisher}{National University
  of Singapore}, \bibinfo{address}{SGP}, \bibinfo{pages}{1–4}.
\newblock


\bibitem[\protect\citeauthoryear{Winograd and Flores}{Winograd and
  Flores}{1986}]%
        {winograd1986understanding}
\bibfield{author}{\bibinfo{person}{Terry Winograd} {and}
  \bibinfo{person}{Fernando Flores}.} \bibinfo{year}{1986}\natexlab{}.
\newblock \bibinfo{booktitle}{\emph{Understanding computers and cognition: A
  new foundation for design}}.
\newblock \bibinfo{publisher}{Intellect Books}.
\newblock


\bibitem[\protect\citeauthoryear{Xing, Wei, Shiratori, and Yatani}{Xing
  et~al\mbox{.}}{2015}]%
        {Xing-autocomplete-anim}
\bibfield{author}{\bibinfo{person}{Jun Xing}, \bibinfo{person}{Li-Yi Wei},
  \bibinfo{person}{Takaaki Shiratori}, {and} \bibinfo{person}{Koji Yatani}.}
  \bibinfo{year}{2015}\natexlab{}.
\newblock \showarticletitle{Autocomplete Hand-Drawn Animations}.
\newblock \bibinfo{journal}{\emph{ACM Trans. Graph.}} \bibinfo{volume}{34},
  \bibinfo{number}{6}, Article \bibinfo{articleno}{169} (\bibinfo{date}{Oct.}
  \bibinfo{year}{2015}), \bibinfo{numpages}{11}~pages.
\newblock
\showISSN{0730-0301}
\urldef\tempurl%
\url{https://doi.org/10.1145/2816795.2818079}
\showDOI{\tempurl}


\bibitem[\protect\citeauthoryear{Yamaoka, Dogan, Bulovic, Saito, Kawahara,
  Kakehi, and Mueller}{Yamaoka et~al\mbox{.}}{2019}]%
        {foldtronics-10.1145/3290605.3300858}
\bibfield{author}{\bibinfo{person}{Junichi Yamaoka},
  \bibinfo{person}{Mustafa~Doga Dogan}, \bibinfo{person}{Katarina Bulovic},
  \bibinfo{person}{Kazuya Saito}, \bibinfo{person}{Yoshihiro Kawahara},
  \bibinfo{person}{Yasuaki Kakehi}, {and} \bibinfo{person}{Stefanie Mueller}.}
  \bibinfo{year}{2019}\natexlab{}.
\newblock \showarticletitle{FoldTronics: Creating 3D Objects with Integrated
  Electronics Using Foldable Honeycomb Structures}. In
  \bibinfo{booktitle}{\emph{Proceedings of the 2019 CHI Conference on Human
  Factors in Computing Systems}} (Glasgow, Scotland Uk)
  \emph{(\bibinfo{series}{CHI '19})}. \bibinfo{publisher}{Association for
  Computing Machinery}, \bibinfo{address}{New York, NY, USA},
  \bibinfo{pages}{1–14}.
\newblock
\showISBNx{9781450359702}
\urldef\tempurl%
\url{https://doi.org/10.1145/3290605.3300858}
\showDOI{\tempurl}


\bibitem[\protect\citeauthoryear{Zaman, Stuerzlinger, Neugebauer, Woodbury,
  Elkhaldi, Shireen, and Terry}{Zaman et~al\mbox{.}}{2015}]%
        {Zaman_Stuerzlinger_Neugebauer_Woodbury_Elkhaldi_Shireen_Terry_2015}
\bibfield{author}{\bibinfo{person}{Loutfouz Zaman}, \bibinfo{person}{Wolfgang
  Stuerzlinger}, \bibinfo{person}{Christian Neugebauer}, \bibinfo{person}{Rob
  Woodbury}, \bibinfo{person}{Maher Elkhaldi}, \bibinfo{person}{Naghmi
  Shireen}, {and} \bibinfo{person}{Michael Terry}.}
  \bibinfo{year}{2015}\natexlab{}.
\newblock \showarticletitle{GEM-NI: A System for Creating and Managing
  Alternatives In Generative Design}. In \bibinfo{booktitle}{\emph{Proceedings
  of the 33rd Annual ACM Conference on Human Factors in Computing Systems - CHI
  ’15}}. \bibinfo{publisher}{ACM Press}, \bibinfo{pages}{1201–1210}.
\newblock
\showISBNx{978-1-4503-3145-6}
\urldef\tempurl%
\url{https://doi.org/10.1145/2702123.2702398}
\showDOI{\tempurl}


\bibitem[\protect\citeauthoryear{Zhao, Kim, Herman, Pfister, Lau, Echevarria,
  and Bylinskii}{Zhao et~al\mbox{.}}{2020}]%
        {Zhao-iconate}
\bibfield{author}{\bibinfo{person}{Nanxuan Zhao}, \bibinfo{person}{Nam~Wook
  Kim}, \bibinfo{person}{Laura~Mariah Herman}, \bibinfo{person}{Hanspeter
  Pfister}, \bibinfo{person}{Rynson~W.H. Lau}, \bibinfo{person}{Jose
  Echevarria}, {and} \bibinfo{person}{Zoya Bylinskii}.}
  \bibinfo{year}{2020}\natexlab{}.
\newblock \showarticletitle{ICONATE: Automatic Compound Icon Generation and
  Ideation}. In \bibinfo{booktitle}{\emph{Proceedings of the 2020 CHI
  Conference on Human Factors in Computing Systems}} (Honolulu, HI, USA)
  \emph{(\bibinfo{series}{CHI '20})}. \bibinfo{publisher}{Association for
  Computing Machinery}, \bibinfo{address}{New York, NY, USA},
  \bibinfo{pages}{1–13}.
\newblock
\showISBNx{9781450367080}
\urldef\tempurl%
\url{https://doi.org/10.1145/3313831.3376618}
\showDOI{\tempurl}


\end{thebibliography}

\appendix

\end{document}